\documentclass[]{interact}

\usepackage{epstopdf}% To incorporate .eps illustrations using PDFLaTeX, etc.
\usepackage[caption=false]{subfig}% Support for small, `sub' figures and tables

\usepackage{amsmath}
\usepackage{amssymb}
\usepackage[version=3]{mhchem}

\theoremstyle{plain}% Theorem-like structures provided by amsthm.sty

\theoremstyle{definition}

\theoremstyle{remark}

\newcommand{\rNc}{N_\mathrm{C}}
\newcommand{\rNo}{N_\mathrm{O}}
\newcommand{\rNv}{N_\mathrm{V}}
\newcommand{\rX}{\mathrm{X}}
\newcommand{\rLDA}{\mathrm{LDA}}
\usepackage[numbers]{natbib}
\begin{document}

%\articletype{ARTICLE TEMPLATE}% Specify the article type or omit as appropriate

\title{Spin-adapted open-shell time-dependent density functional theory:
towards a simple and accurate method for spin-flip-down excitations }

\author{
\name{Hewang Zhao and Zhendong Li\thanks{ Email: zhendongli@bnu.edu.cn} }
\affil{Key Laboratory of Theoretical and Computational Photochemistry, Ministry of Education, College of Chemistry, Beijing Normal University, Beijing 100875, China}
}

\maketitle

\begin{abstract}
A major challenge in using spin-flip time-dependent density functional theory (SF-TD-DFT) for spin-flip-down excitations
is the presence of spin contamination. While several improved methods have been developed in the past,
a simple and accurate method remains elusive.
Here, based on our previous development on spin-adapted open-shell TD-DFT for spin-conserving excitations (X-TD-DFT) [Z. Li and W. Liu, J. Chem. Phys. 135, 194106 (2011)], we introduce a method termed as XSF-TDA for modeling spin-flip-down excitations, and provide an in-depth comparison of different methodologies for mitigating spin contamination in SF-TDA. Pilot applications to prototype systems demonstrate the promise of XSF-TDA over existing SF-TDA methods, including unrestricted SF-TDA (USF-TDA) and mixed-reference SF-TDA (MRSF-TDA),
in describing bond breakings and inverted singlet-triplet gap systems.
\end{abstract}

\begin{keywords}
spin-flip; time-dependent density functional theory; spin contamination; spin adaptation
\end{keywords}

\section{Introduction}

Spin-flip time-dependent density functional theory (SF-TD-DFT)\cite{shao2003spin}, which extends the earlier spin-flip framework\cite{casanova2020spin} to TD-DFT, 
has garnered significant attention for its ability to describe electronic transitions beyond the reach of conventional linear response spin-conserving TD-DFT. For instance, it can be used to study spin-flip transitions in spin-flip emitters\cite{kitzmann2022spin,kitzmann2023charge}, which represent a distinct class of luminescent materials that exhibit unique photophysical properties due to spin‐forbidden yet radiative transitions. 
Apart from spin-flip transitions, SF-TD-DFT is also capable of capturing certain double excitations, single-bond dissociation processes,
and the correct topology of conical intersections\cite{zhang2015spin,2019Analytic,Keipert2014Dynamics}, which are notoriously challenging for DFT and spin-conserving TD-DFT within the adiabatic approximation\cite{herbert2022spin}.

The initial implementation of SF-TD-DFT by Shao et al.\cite{shao2003spin} employed a collinear hybrid exchange–correlation (XC) kernel within the Tamm–Dancoff approximation (TDA). This formulation necessitates the use of hybrid XC functionals; pure density functionals would otherwise yield excitation energies that are simply orbital energy differences. Subsequently, Wang and Ziegler introduced a noncollinear XC kernel\cite{wang2004time,wang2005performance},
which was originally proposed in the context of relativistic TD-DFT for spinor excitations\cite{wang2003comparison,gao2004time,gao2005time},
within the local density approximation (LDA)
to describe spin-flip excitations in a nonrelativistic framework. 
Later, full noncollinear XC kernels were implemented for generalized gradient approximation (GGA) and hybrid functionals, both within TDA\cite{rinkevicius2010spin,bernard2012general} and 
in full SF-TD-DFT\cite{li2012theoretical}. 
However, numerical instabilities can arise with noncollinear GGA kernels. To address this, an adiabatic local density approximation (ALDA0)\cite{li2012theoretical,li2016critical} was introduced, neglecting density gradients in the XC kernel evaluation to enhance numerical stability. More recently, alternative strategies based on a multicollinear formulation have been proposed\cite{li2023noncollinear,zhang2025spin,wang2025zero}.

\begin{figure}
    \centering
    \includegraphics[width=0.48\textwidth]{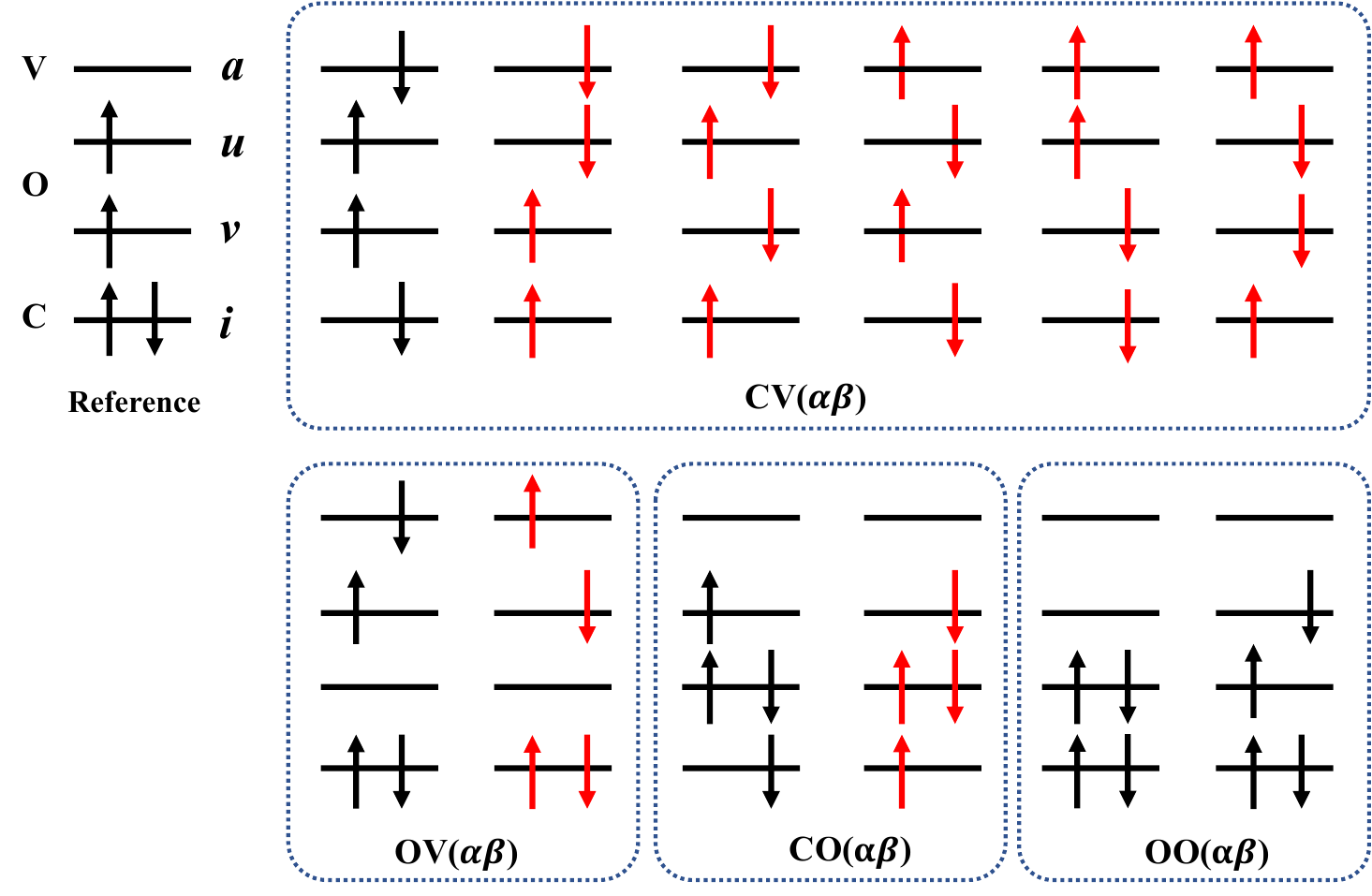}
    \caption{\raggedright 
    Illustration of the excited-state spin contamination problem in SF-TD-DFT/SF-TDA
    due to the incompleteness of the excited state manifold generated by spin-flip-down single excitations. Configurations in red color represent the missing higher excitations with respect to the reference
    to achieve spin completeness/adaptation.}
    \label{fig:confs}
\end{figure}

A major drawback of SF-TD-DFT is the presence of excited-state spin contamination\cite{casida2005propagator,li2010Spin}, which arises from the incompleteness of the excitation manifold. We define a subspace as \emph{spin complete} if it contains all the determinants required to form spin eigenfunctions, and a method as \emph{spin-adapted} if all its configurations are spin eigenfunctions. Note a subtle distinction: spin-adaptation requires that only spin eigenfunctions are included, but does not necessitate inclusion of all the configurations within a spin-complete subspace.
To illustrate the spin contamination issue in SF-TD-DFT, consider a triplet reference system as depicted in Fig. \ref{fig:confs}. The orbitals of the high-spin open-shell state can be categorized into closed-shell (C), open-shell (O), and virtual-shell (V). We denote a single excitation from an $\alpha$ core orbital to a $\beta$ virtual orbital as CV($\alpha\beta$). It is easy to find that only OO($\alpha\beta$)-type spin-flip-down excitations generate a spin-complete subspace. In contrast, the other three types - CV($\alpha\beta$), OV($\alpha\beta$), and CO($\alpha\beta$) - result in spin contamination due to the spin incompleteness of their generated configurations.
The determinants shown in red in Fig \ref{fig:confs},
which are higher excitations than singles
with respect to the reference,
are necessary to achieve spin completeness/adaptation.

Within wavefunction theory, methods such as spin-complete spin-flip configuration interaction singles (SC-SF-CIS)\cite{sears2003spin} have been developed to address this issue by directly incorporating the missing higher excitations. However, adapting such approaches to TD-DFT is challenging due to the absence of an explicit wavefunction. Although propagator corrections\cite{casida2005propagator} and generalized excitation operators\cite{vahtras2007general} have been proposed to include higher-order excitations in TD-DFT, these methods have not seen practical application.

Building on the insight that the missing configurations for spin completeness can be accessed from other components of the open-shell reference, one of the present authors developed\cite{li2010Spin} a spin-adapted random phase approximation (S-RPA) and time-dependent density functional theory (S-TD-DFT) using the tensor equation-of-motion (TEOM) formalism originally introduced in nuclear physics\cite{rowe1975tensor}. A key feature of S-RPA is that only single excitations explicitly enter the final matrix elements, while the contributions from higher excitations required for spin-adaptation are implicitly incorporated via the Wigner-Eckart theorem\cite{li2010Spin}. This crucial property allows for a direct translation of the matrix elements into the TD-DFT framework, leading to S-TD-DFT\cite{li2011spin} without the need to go beyond the adiabatic approximation. 
Subsequently, a more accurate and simpler version, termed X-TD-DFT\cite{li2011spin2}, was introduced to better account for spin degeneracy conditions and to eliminate double counting of correlation for those already spin-adapted states. 
X-TD-DFT has been shown to achieve accuracy comparable to conventional closed-shell TD-DFT for spin-conserving excitations\cite{li2016criticalDD}.

For spin-flip-down excitations, a similar strategy can in principle be applied to achieve spin adaptation, as 
proposed in the original work\cite{li2010Spin} and
later implemented by Zhang and Herbert\cite{zhang2015spin} and Chima and Visscher\cite{chibueze2025spin}. 
However, addressing double counting of correlation in this context is more challenging, as spin contamination affects three types of excitations, and adding the effects of higher excitations for spin-adaptation may deteriorate the excited states dominated by the OO($\alpha\beta$) type of excitations, which are already spin-complete.
Another relevant approach is the mixed-reference SF-TD-DFT within TDA (MRSF-TDA) proposed by Lee et al.\cite{lee2018eliminating,lee2019efficient}, which uses only the $M_S = \pm 1$ components of a triplet reference to mitigate spin contamination, but is not applicable to other spin states. In summary, a general, simple, and accurate spin-adapted method for spin-flip-down excitations is still lacking.

In this work, we present a method termed as XSF-TDA, which extends our previous developments in spin-adapted TD-DFT\cite{li2010Spin,li2011spin,li2011spin2} to spin-flip-down excitations. We introduce a practical approach to alleviate the double counting of correlation resulting from the implicit inclusion of high-order excitations for spin adaptation. Pilot applications on small systems, where high-level theoretical or experimental reference data are available, demonstrate that XSF-TDA performs comparably to unrestricted SF-TDA (USF-TDA) for excited states without significant spin contamination, and consistently outperforms USF-TDA for states exhibiting strong spin contamination. Formal and numerical comparisons are also made between XSF-TDA and MRSF-TDA.

The remainder of this paper is organized as follows. In Sec. \ref{sec:theory}, we first briefly review the SF-TDA formalism and introduce tensor configuration interaction singles (TCIS) based on the TEOM framework. We then derive XSF-TDA from TCIS and discuss its relationship to existing approaches. Section~\ref{sec:results} presents benchmark results obtained with USF-TDA, MRSF-TDA, and XSF-TDA. Conclusions and prospects for future developments are provided in the last section. The following convention for labeling molecular orbitals is used throughout the paper: $\{i,j,k,l,\cdots\}$ for doubly occupied, $\{t,u,v,w,\cdots\}$
for singly occupied, $\{a,b,c,d,\cdots\}$ for unoccupied, and $\{p,q,r,s,\cdots\}$ for unspecified orbitals. Greek indices (e.g., $\alpha$ and $\beta$) are used to denote electron spin, and the notation $\bar{p}$ refers to the $\beta$ spin-orbital corresponding to the $\alpha$
spin-orbital $p$.

\section{Theory}\label{sec:theory}
\subsection{Spin-flip TD-DFT within the Tamm-Dancoff approximation (SF-TDA)}\label{sec:SFTDA}
The SF-TDA for spin-flip-down excitations 
amounts to solving the following eigenvalue problem
\begin{equation}
\mathbf{A}^{\mathrm{SF\textrm{-}TDA}}\mathbf{X}_\lambda = \omega_\lambda \mathbf{X}_\lambda,
\label{eq:SFTDA}
\end{equation}
where the matrix elements of $\mathbf{A}^{\mathrm{SF\textrm{-}TDA}}$ are given by
\begin{equation}
A_{\bar{p}q,\bar{r}s}^{\mathrm{SF\textrm{-}TDA}} = \delta_{qs} f_{pr}^\beta - \delta_{pr} f_{sq}^\alpha + K_{\bar{p}q,\bar{r}s}.
\label{eq:SFTDA_amat}
\end{equation}
The Kohn–Sham matrix $f^{\sigma}$ for spin $\sigma$ is generally expressed as
\begin{eqnarray}
    f^{\sigma}_{pq}=&h^{\sigma}_{pq}+\sum_{i\sigma '}(p_{\sigma}q_{\sigma}|i_{\sigma^{'}}i_{\sigma '})+(p_{\sigma}|v_{XC}^{\sigma}|q_{\sigma})\nonumber\\
    &-\sum_{i\sigma '}c_{\rX} (p_{\sigma}i_{\sigma '}|i_{\sigma '}q_{\sigma}),\label{eq:ks_fock}
\end{eqnarray}
which is applicable to LDA, GGA, and hybrid functionals. Here, $h^{\sigma}$ denotes the one-electron term including kinetic energy and nuclear attraction; the second term represents the Coulomb repulsion in Mulliken notation; $v_{\text{XC}}^{\sigma}$ is the exchange–correlation (XC) potential for spin $\sigma$; and the last term introduces exact exchange, with $c_{\rX}$ denoting the fraction of exact exchange in hybrid functionals.
The coupling matrix elements $K_{\bar{p}q,\bar{r}s}$ take the general form
\begin{equation}
K_{\bar{p}q,\bar{r}s} = 
- c_\rX (\bar{p} \bar{r} | s q)
+ [K_{\text{XC}}]_{\bar{p}q,\bar{r}s},
\end{equation}
where the exchange–correlation kernel $[K_{\text{XC}}]_{\bar{p}q,\bar{r}s}$ for spin-flip excitations depends on the parameterization of the XC functional. While this term vanishes for collinear functionals, it is expressed for noncollinear functionals as\cite{wang2004time}
{\small
\begin{align}
[K_{\text{XC}}]_{\bar{p}q,\bar{r}s} =
\int \psi_{\bar{p}}(\mathbf{r}) \psi_{q}(\mathbf{r})
\left[ \frac{ v_{\text{XC}}^{\alpha}(\mathbf{r}) - v_{\text{XC}}^{\beta}(\mathbf{r}) }{ \rho_{\alpha}(\mathbf{r}) - \rho_{\beta}(\mathbf{r}) } \right]
\psi_{s}(\mathbf{r}) \psi_{\bar{r}}(\mathbf{r}) d\mathbf{r}.
\label{eq:SFkernel}
\end{align}}Although noncollinear functionals are theoretically more general than their collinear counterparts, the evaluation of Eq. \eqref{eq:SFkernel} for GGA-type functionals can encounter numerical instabilities in regions where $\rho_{\alpha}(\mathbf{r}) \approx \rho_{\beta}(\mathbf{r})$ but $\nabla\rho_{\alpha}(\mathbf{r}) \ne \nabla\rho_{\beta}(\mathbf{r})$. To mitigate this issue, the ALDA0 kernel\cite{li2012theoretical} is introduced, which effectively neglects density gradient contributions in Eq. \eqref{eq:SFkernel} by setting $\nabla\rho_\sigma = 0$
in computing $v_{\text{XC}}^\sigma$. In this work, we adopt the ALDA0 kernel for GGA and hybrid functionals. The application of multicollinear functionals\cite{li2023noncollinear,zhang2025spin} will be explored in future studies.
    
%It is seen from Eq.\ref{A_B_mats} that the TDA flip-up and flip-down excitations are decopled, means that they can be treated separately. Compared with the full TD-DFT, TDA is not only very accurate but also is immune to the so-called(near) instability problem due to the decoupling of excitations and de-excitations.  In particular, when the calculated excitation energies are rather small, the energy gradients of TDA remain finite but those of TD-DFT may approach infinity.\cite{russ2004real} Therefore, SF-TDA is particularly suited for studying global potential energy surfaces.}\\

\subsection{Tensor configuration interaction singles (TCIS)}
We mainly focus on TDA in this work, and will use a CIS analog of TEOM, referred to as tensor configuration interaction singles (TCIS). Another important motivation for this choice is that TEOM will lead to non-Hermitian equations for spin-flip-down excitations. To illustrate their differences, we first reexamine the conventional CIS and EOM. Consider the
following CI-like equation for an excitation operator $\hat{O}_\lambda^\dagger$ for the $\lambda$-th excited state,
\begin{eqnarray}
(\hat{H}-E_0)\hat{O}_\lambda^\dagger|0\rangle = \omega_\lambda \hat{O}_\lambda^\dagger|0\rangle,\label{eq:CIS}
\end{eqnarray}
where $|0\rangle$ denotes the reference, $E_0=\langle 0|\hat{H}|0\rangle$ is the reference energy,
and $\omega_\lambda$ represents the excitation energy.
When $\hat{O}_\lambda^\dagger = \sum_{J} q_J^\dagger X_{J,\lambda}$ is expanded in an operator basis $\{q_J^\dagger\}$, the expansion coefficients $X_{J,\lambda}$ can be determined by solving an Hermitian eigenvalue problem
\begin{eqnarray}
\sum_{J}
\langle 0|q_I (\hat{H}-E_0) q_J^\dagger|0\rangle X_{J,\lambda}
= \omega_\lambda
\sum_{J}
\langle 0|q_I q_J^\dagger|0\rangle X_{J,\lambda},\label{eq:CIS2}
\end{eqnarray}
It differs from the EOM formulation (of the first kind\cite{li2010Spin})
\begin{eqnarray}
\sum_{J}
\langle 0|q_I [\hat{H},q_J^\dagger]|0\rangle X_{J,\lambda}
= \omega_\lambda
\sum_{J}
\langle 0|[q_I,q_J^\dagger]|0\rangle X_{J,\lambda},\label{eq:EOM}
\end{eqnarray}
in two aspects. First, if the de-excitation operator $q_I$ satisfies
the 'killer condition' $q_I|0\rangle=0$, then the right-hand sides (RHS) of Eqs. \eqref{eq:CIS2} and \eqref{eq:EOM} become identical. Second, the left-hand 
sides (LHS) of Eqs. \eqref{eq:CIS2} and \eqref{eq:EOM} coincide if the condition 
$\langle 0|q_Iq_J^\dagger|0\rangle E_0 = \langle 0|q_Iq_J^\dagger \hat{H}|0\rangle$ holds.
While these two conditions are satisfied in standard CIS and EOM treatments of closed-shell systems,
they may not hold in the case of TCIS and TEOM for spin-flip-down excitations in open-shell systems, as will be discussed later in Sec. \ref{sec:comparison}.

To obtain TCIS for spin-adaptation of spin-flip-down excitations, 
the single excitation operators are recombined to form a 
triplet excitation operator $T^{\dagger}_{pq}(1)$ defined as follows\cite{1981Second}
\begin{eqnarray}
  T^{\dagger}_{pq}(1,1) &=&-p^{\dagger}_{\alpha}q_{\beta},\\
  T^{\dagger}_{pq}(1,0) &=& \frac{1}{\sqrt{2}}(p^{\dagger}_{\alpha}q_{\alpha}-p^{\dagger}_{\beta}q_{\beta}),\\
  T^{\dagger}_{pq}(1,-1) &=& p^{\dagger}_{\beta}q_{\alpha},
\end{eqnarray}
where $p^{(\dagger)}_{\alpha}$ denotes the annihilation (creation) operator.
A tensor reference $|S_{i}\rangle\rangle$ is defined as the collection of all components of a spin multiplet $\{|S_{i}M_{i}\rangle :\, M_{i}=-S_{i},\dots, S_{i}\}$ 
with initial spin $S_i$.
The excited configuration $|S_{f}\rangle\rangle_{pq}$ resulting from the actions of $T^{\dagger}_{pq}(1)$ on $|S_{i}\rangle\rangle$ is expressed as 
\begin{eqnarray}
  |S_{f}\rangle\rangle_{pq}=[T^{\dagger}_{pq}(1)\times|S_{i}\rangle\rangle]^{S_{f}},\label{eq:TCISconfs}
\end{eqnarray}
where $S_f=S_i-1$ is considered in this work. 
For spin-flip-down single excitations,
the excited configurations can be generated by
CV(1), CO(1), OV(1), and OO(1) type of excitations, 
where the symbol CV(1) denotes operators
of form $T_{ai}^\dagger(1)$. 
For the OO(1) type of excitations, it is
useful to distinguish cases where $t\ne u$ 
from those where $t=u$ in $T_{tu}^\dagger(1)$.
We adopt the notation O$_1$O$_2$(1) for $t\ne u$ and O$_1$O$_1$(1) for $t=u$, respectively.
The complete operator basis for excited states with spin $S_f=S_i-1$ is thus given by
\begin{align}
\{q_J^\dagger\}=
\{T^{\dagger}_{
\mathrm{CV}}(1),
T^{\dagger}_{
\mathrm{CO}}(1),
T^{\dagger}_{
\mathrm{OV}}(1),
T^{\dagger}_{\mathrm{O_{1}O_{2}}}(1),
T^{\dagger}_{\mathrm{O_{1}O_{1}}}(1)\}.\label{eq:OpBasis}
\end{align}

The spin-adapted counterpart of Eq. \eqref{eq:CIS2}, referred to as TCIS, can be obtained
in the same way as TEOM\cite{li2010Spin}
\begin{eqnarray}
\mathbf{M}_{S_{f}}\mathbf{X}_{\lambda}&=&\omega_{\lambda} \mathbf{N}_{S_{f}}\mathbf{X}_{\lambda},\label{eq:TCIS}
\end{eqnarray}
where both $\mathbf{M}_{S_{f}}$ and $\mathbf{N}_{S_{f}}$
are Hermitian. The overlap matrix $\mathbf{N}_{S_{f}}$ corresponding to the operator basis defined in Eq. \eqref{eq:OpBasis} 
is block-diagonal
\begin{eqnarray}
\mathbf{N}_{S_{f}} &=&
\mathbf{N}_{S_{f}}^{\mathrm{CV}}\oplus
\mathbf{N}_{S_{f}}^{\mathrm{CO}}\oplus
\mathbf{N}_{S_{f}}^{\mathrm{OV}}\oplus
\mathbf{N}_{S_{f}}^{\mathrm{O_1O_2}}\oplus
\mathbf{N}_{S_{f}}^{\mathrm{O_1O_1}}, \nonumber\\
\protect[\mathbf{N}_{S_{f}}^{\mathrm{CV}}]_{ai,bj} &=& \delta_{ij}\delta_{ab},\nonumber\\
\protect[\mathbf{N}_{S_{f}}^{\mathrm{CO}}]_{ui,vj} &=& \frac{2S_i+1}{2S_i}\delta_{ij}\delta_{uv},\nonumber\\
\protect[\mathbf{N}_{S_{f}}^{\mathrm{OV}}]_{au,bv} &=& \frac{2S_i+1}{2S_i}\delta_{uv}\delta_{ab},\nonumber\\
\protect[\mathbf{N}_{S_{f}}^{\mathrm{O_1O_2}}]_{tu,vw} &=& \frac{2S_i+1}{2S_i-1}\delta_{tv}\delta_{uw},\quad t\ne u,\, v\ne w,\nonumber\\
\protect[\mathbf{N}_{S_{f}}^{\mathrm{O_1O_1}}]_{tt,vv} &=& 
\frac{2S_i+1}{2S_i-1}(\delta_{tv}-\frac{1}{2S_i}).
\end{eqnarray}
These expressions indicate that the configurations defined by Eqs. \eqref{eq:TCISconfs} and
\eqref{eq:OpBasis} are generally non-orthonormal. We can first apply 
the following scaling transformation
\begin{eqnarray}
\mathbf{W} &=& \mathbf{I}^{\mathrm{CV}} \oplus
\sqrt{\frac{2S_i}{2S_i+1}}\mathbf{I}^{\mathrm{CO}} \oplus
\sqrt{\frac{2S_i}{2S_i+1}}\mathbf{I}^{\mathrm{OV}} \oplus \nonumber\\
&&
\sqrt{\frac{2S_i-1}{2S_i+1}}\mathbf{I}^{\mathrm{O_1O_2}} \oplus
\sqrt{\frac{2S_i-1}{2S_i+1}}\mathbf{I}^{\mathrm{O_1O_1}},\label{eq:Wtrans}
\end{eqnarray}
such that all configurations become orthonormal except for the O$_1$O$_1$ type. The metric for this type
of excitations is of rank $\rNo-1$,
where $\rNo=2S_i$ denotes the number of open-shell orbitals, because their equal superposition simply yields another component of the reference, viz.,
\begin{eqnarray}
\sum_{t}T^\dagger_{tt}(1,-1)|S_iS_i\rangle
=\hat{S}_-|S_iS_i\rangle \propto    
|S_i(S_i-1)\rangle.
\end{eqnarray}
An orthonormal set of spin-adapted basis for $S_f=S_i-1$ can be generated by an isometry
$\mathbf{V}=\mathbf{I}^{\mathrm{CV}}\oplus \mathbf{I}^{\mathrm{CO}}\oplus \mathbf{I}^{\mathrm{OV}} \oplus\mathbf{I}^{\mathrm{O_1O_2}}\oplus\mathbf{V}^{\mathrm{O_1O_1}}$, whose nontrivial block 
$\mathbf{V}^{\mathrm{O_1O_1}}$ is
an $\rNo\times (\rNo-1)$ matrix
that can be chosen as\cite{1979Spin}
\begin{equation}
    \mathbf{V}^{\mathrm{O_1O_1}}=\begin{bmatrix}
        \frac{\rNo-1}{\sqrt{\rNo(\rNo-1)}}& 0 & \dots & 0\\
			-\frac{1}{\sqrt{\rNo(\rNo-1)}}& \frac{\rNo-2}{\sqrt{(\rNo-1)(\rNo-2)}} & \dots&0\\
			\vdots&\vdots&\ddots&\vdots\\
			-\frac{1}{\sqrt{\rNo(\rNo-1)}}& -\frac{1}{\sqrt{(\rNo-1)(\rNo-2)}}&\dots&\frac{1}{\sqrt{2}}\\
			-\frac{1}{\sqrt{\rNo(\rNo-1)}}&-\frac{1}{\sqrt{(\rNo-1)(\rNo-2)}}&\dots&-\frac{1}{\sqrt{2}}
    \end{bmatrix},\label{eq:Vo1o1}
\end{equation}
where the equal superposition is eliminated.
A summary of the final orthonormalized excited configurations is provided in Table \ref{tab:orthoconfs}.

\begin{table*}[htbp]
  \centering
  \caption{\raggedright Orthonormalized excited configurations for
  $S_{f}=S_{i}-1$. The matrix elements $V_{t,k}$ ($k=1,\cdots,\rNo-1$) are given in Eq. \eqref{eq:Vo1o1}. $\rNc$, $\rNo$, and $\rNv$ refer to the number of close-shell, open-shell, vacant-shell orbitals, respectively.}
  \scalebox{0.8}{
 \begin{tabular}{cccc}
     \hline\hline
     type & dim & excited configuration & expanded form for the high-spin component \\
     \hline
      CV(1) & $\rNc\rNv$ & 
      $[T_{ai}^\dagger\times|S_i\rangle\rangle]^{S_f}$ & 
      $\sqrt{\frac{2S_{i}-1}{2S_{i}+1}}\left(|\Psi_{i}^{\bar a}\rangle 
      -\frac{1}{2S_{i}}\sum_{t}(
      |\Psi_{it}^{a\bar t}\rangle -
      |\Psi_{\bar it}^{\bar a\bar t}\rangle)\right)$\\ &&&$
      -\frac{1}{2S_{i}\sqrt{(2S_{i}+1)(2S_{i}-1)}}\sum_{tu}|\Psi_{\bar iut}^{a\bar u\bar t}\rangle$ \\
     CO(1) & $\rNc\rNo$ & $\sqrt{\frac{2S_i}{2S_i+1}} [T_{ui}^\dagger\times|S_i\rangle\rangle]^{S_f}$ &$|\Psi_{i}^{\bar u}\rangle\sqrt{\frac{2S_{i}-1}{2S_{i}}}+\sum_{t}|\Psi_{\bar it}^{\bar u\bar t}\rangle\frac{1}{\sqrt{2S_{i}(2S_{i}-1)}}$\\
     OV(1) & $\rNo\rNv$ & $\sqrt{\frac{2S_i}{2S_i+1}}[T_{au}^\dagger\times|S_i\rangle\rangle]^{S_f}$ & $|\Psi_{u}^{\bar a}\rangle\sqrt{\frac{2S_{i}-1}{2S_{i}}}-\sum_{t}|\Psi_{ut}^{a\bar t}\rangle\frac{1}{\sqrt{2S_{i}(2S_{i}-1)}}$ \\
     O$_1$O$_2$(1) & $\rNo(\rNo-1)$ & 
     $\sqrt{\frac{2S_i-1}{2S_i+1}}[T_{tu}^\dagger\times|S_i\rangle\rangle]^{S_f}$ & $|\Psi^{\bar{t}}_u\rangle$ ($t\ne u$) \\
     O$_1$O$_1$(1) & $\rNo-1$ & $\sqrt{\frac{2S_i-1}{2S_i+1}}\sum_{t}[T_{tt}^\dagger\times|S_i\rangle]^{S_f} V_{t,k}$ & 
     $\sum_{t}(|\Psi_{t}^{\bar t}\rangle-
     \frac{1}{2S_{i}}\sum_{v}|\Psi_{v}^{\bar v}\rangle)V_{t,k}=
     \sum_{t} |\Psi_{t}^{\bar t}\rangle V_{t,k}$\\
     \hline\hline
 \end{tabular} }
 \label{tab:orthoconfs}
\end{table*}

Using Eqs. \eqref{eq:Wtrans} and \eqref{eq:Vo1o1}, we can transform Eq. \eqref{eq:TCIS} into a standard eigenvalue problem
\begin{eqnarray}
\tilde{\mathbf{A}}^{\mathrm{TCIS}}\tilde{\mathbf{X}}_\lambda =\omega_\lambda \tilde{\mathbf{X}}_\lambda,
\end{eqnarray}
where $\tilde{\mathbf{A}}^{\mathrm{TCIS}}$ is defined by
\begin{eqnarray}
\tilde{\mathbf{A}}^{\mathrm{TCIS}}=\mathbf{V}^T \mathbf{A}(S_i) \mathbf{V},\quad
\mathbf{A}(S_i) = \mathbf{W}^T\mathbf{M}\mathbf{W}.\label{eq:AmatTCIS}
\end{eqnarray}
Here, the subscript $S_f$ has been omitted for brevity, but the dependence on $S_i$ in $\mathbf{A}(S_i)$ is explicitly indicated.
Following the approach detailed in Ref. \cite{li2010Spin}, we can find all matrix elements
of $\mathbf{A}(S_i)$, which are summarized in Table \ref{tab:tcis}. 
It is important to note that the resulting matrix $\mathbf{A}(S_i)$
is not equivalent to that of SC-SF-CIS\cite{sears2003spin}, as its dimension
is exactly the same as that of SF-CIS,
while SC-SF-CIS contains more configurations for
spin-completeness.
For instance, considering the CV($\alpha\beta$) type of excitations illustrated in Fig. \ref{fig:confs}, TCIS only contains one spin-adapted configuration generated from Eq. \eqref{eq:TCISconfs}, whereas SC-SF-CIS includes all the two spin-adapted singlet configurations
from the the six underlying determinants. A key advantage of TCIS lies in that 
its ability to seamlessly bridge the spin-contaminated SF-CIS with a spin-adapted formulation. 
Specifically, taking the limit $S_i\rightarrow\infty$
in $\mathbf{A}(S_i)$ recovers the conventional SF-CIS matrix elements $\mathbf{A}(\infty)$.
This property enables a natural extension of our previously developed X-TD-DFT framework for spin-conserving excitations\cite{li2011spin2}, facilitating the construction of the XSF-TDA method for 
spin-flip-down excitations.

\begin{table*}[htbp]
    \centering
    \caption{\raggedright Matrix elements of $\mathbf{A}(S_i)$ in Eq. \eqref{eq:AmatTCIS} for TCIS. Only the upper triangular parts are presented as the matrix is Hermitian. In the limit of $S_i\rightarrow\infty$, all the matrix elements go to those in SF-CIS,
    i.e., Eq. \eqref{eq:SFTDA_amat} with $c_\rX=1$. 
    %{\color{red}matrix elements of TCIS for abitrary orbitals?}
    The matrix elements $f_{pq}^{S}$ is defined by
    $f_{pq}^{S}=\frac{1}{2}(f_{pq}^{\beta}-f_{pq}^{\alpha})=\frac{1}{2}\sum_t (pt|tq)$.
    }
    \begin{tabular}{cl}
    \hline
    \hline
         block &  matrix elements\\
         \hline
         CV(1)-CV(1) & $A_{ai,bj}=\delta_{ij}f_{ab}^{\beta}-\delta_{ab}f_{ji}^{\alpha}+\frac{1}{S_{i}}\delta_{ij}f_{ab}^{S}+\frac{1}{S_{i}}\delta_{ab}f_{ji}^{S}-(ab|ji)$\\
         CO(1)-CO(1) & $A_{ui,vj}=\delta_{ij}f_{uv}^{\beta}-\delta_{uv}f_{ji}^{\alpha}+\frac{2}{2S_{i}-1}\delta_{uv}f_{ji}^{S}-(uv|ji)-\frac{1}{2S_{i}-1}(ui|jv)$\\
         OV(1)-OV(1) & $A_{au, bv}=\delta_{uv}f_{ab}^{\beta}-\delta_{ab}f_{uv}^{\alpha}+\frac{2}{2S_{i}-1}\delta_{uv}f_{ab}^{S}-(ab|vu)-\frac{1}{2S_{i}-1}(au|vb)$\\
         CV(1)-CO(1) & $A_{ai,vj}=\sqrt{\frac{2S_{i}+1}{2S_{i}}}\delta_{ij}f_{av}^{\beta}-\sqrt{\frac{2S_{i}+1}{2S_{i}}}(av|ji)$\\
         CV(1)-OV(1) & $A_{ai,bv}=-\sqrt{\frac{2S_{i}+1}{2S_{i}}}\delta_{ab}f_{vi}^{\alpha}-\sqrt{\frac{2S_{i}+1}{2S_{i}}}(ab|vi)$\\
         CO(1)-OV(1) & $A_{ui,bv}=-\frac{2S_{i}}{2S_{i}-1}(ub|vi)+\frac{1}{2S_{i}-1}(ui|vb)$\\
         OO(1)-OO(1) & $A_{tu,vw}=\delta_{wu}f_{tv}^{\beta}-\delta_{tv}f_{uw}^{\alpha}-(tv|wu)$\\
         CV(1)-OO(1) & $A_{ai,vw}=\sqrt{\frac{2S_{i}+1}{2S_{i}-1}}(\delta_{vw}\frac{1}{S_{i}}f_{ia}^{s}-(av|wi))$\\
         CO(1)-OO(1) & $A_{ui,vw}=\sqrt{\frac{2S_{i}}{2S_{i}-1}}(-\delta_{vu}f_{iw}^{\alpha}-(uv|wi))+\frac{1}{\sqrt{2S_{i}(2S_{i}-1)}}\delta_{vw}f_{iu}^{\beta}$\\
         OV(1)-OO(1) & $A_{au,vw}=\sqrt{\frac{2S_{i}}{2S_{i}-1}}(\delta_{wu}f_{av}^{\beta}-(av|wu))-\frac{1}{\sqrt{2S_{i}(2S_{i}-1)}}\delta_{vw}f_{au}^{\alpha}$\\
         \hline
         \hline
    \end{tabular}
    \label{tab:tcis}
\end{table*}

\subsection{Spin-adapted TDA for spin-flip-down excitations (XSF-TDA)}
For spin-conserving excitations, X-TD-DFT\cite{li2011spin2} is a simpler and more accurate alternative than S-TD-DFT obtained simply by a translation rule\cite{li2011spin}, since it takes into account of spin degeneracy conditions and eliminates double counting of correlation for those already spin-adapted states. Motivated by its success, 
we extend a similar strategy to spin-flip–down excitations.
To this end, we express $\mathbf{A}(S_i)$ as 
\begin{eqnarray}
\mathbf{A}(S_i) = \mathbf{A}(\infty) + \Delta\mathbf{A}^{\mathrm{TCIS}}(S_i).
\end{eqnarray}
Since $\mathbf{A}(\infty)$ is just SF-CIS, $\Delta\mathbf{A}^{\mathrm{TCIS}}(S_i)$ can be viewed
as a spin adaptation correction, whose matrix elements
are provided in Table \ref{tab:mats_dA}.
Note that $\Delta\mathbf{A}^{\mathrm{TCIS}}(S_i)=0$
for the OO(1) type of excitations, consistent with the spin-complete nature of this excitation class, as previously discussed.

Following the X-TD-DFT framework\cite{li2011spin2},
we define the matrix $\tilde{\mathbf{A}}^{\mathrm{XSF\textrm{-}TDA}}$ in the XSF-TDA eigenvalue equation by
\begin{eqnarray}
\tilde{\mathbf{A}}^{\mathrm{XSF\textrm{-}TDA}} = \mathbf{V}^T[\mathbf{A}^{\mathrm{SF\textrm{-}TDA}} + g_\rX\Delta\mathbf{A}^{\mathrm{TCIS}}(S_i)]\mathbf{V},\label{eq:XSFTDA}
\end{eqnarray}
where $\mathbf{A}^{\mathrm{SF\textrm{-}TDA}}$ is given by Eq. \eqref{eq:SFTDA_amat} for SF-TDA, and 
$\Delta\mathbf{A}^{\mathrm{TCIS}}(S_i)$ is the spin adaptation correction derived from TCIS, evaluated using orbitals from restricted open-shell Kohan-Sham (ROKS) calculations. In contrast to X-TD-DFT, which mitigates double counting of correlation simply by removing specific excitation couplings\cite{li2011spin2}, we introduce a global hybridization factor $g_\rX(c_\rX)$, which is a function of the fraction of Hartree-Fock exchange $c_\rX$. For the Hartree-Fock case ($c_\rX=1$), we require $g_\rX(1)=1$ such that Eq. \eqref{eq:XSFTDA} becomes identical to TCIS. For the case with XC functionals ($c_\rX<1$), we require $0<g_\rX(c_\rX)<1$, such that the double counting of correlation introduced by $\Delta\mathbf{A}^{\mathrm{TCIS}}$ is reduced,
meanwhile, non-trivial corrections for spin-contaminated excitations are still present.
In this work, we adopted a simple linear function form
\begin{eqnarray}
g_\rX(c_\rX)=(1-g_{\rLDA})c_{\rX}+g_{\rLDA},\label{eq:rX} 
\end{eqnarray}
which contains one empirical parameter $g_{\rLDA}$ to be fixed. We use $g_{\rLDA}=0.3$ throughout this work, which is fixed by enforcing the desired degeneracy of the $1^1P_{x,y,z}$ states in the calculation for the Be atom (see Sec. \ref{sec:BeAndMg}) with the aug-cc-pVTZ basis set. 
Extending XSF-TDA to more general XC functionals such as range-separated hybrid functionals and exploring more complex function form other than \eqref{eq:rX}
will be explored in future.

\begin{table*}[t]
		\centering
		\caption{\raggedright Matrix elements of $\Delta\mathbf{A}^{\mathrm{TCIS}}(S_i)$ for spin-down excitations, where only the upper triangular parts are presented. $S_{i}$ is the total spin of the ground state, and the matrix elements $f_{pq}^{S}=\frac{1}{2}(f_{pq}^{\beta}-f_{pq}^{\alpha})=\frac{1}{2}\sum_t (pt|tq)$.}\label{tab:mats_dA}
		\begin{tabular}{cl}
			\hline
			\hline
			Block & Matrix elements  \\
			\hline
			CV(1)-CV(1) & $\Delta A_{ai,bj}=\frac{1}{S_{i}}(\delta_{ij}f_{ab}^{S}+\delta_{ab}f_{ji}^{S})$ \\
			CO(1)-CO(1) & $\Delta A_{ui,vj}=\frac{2}{2S_{i}-1}\delta_{uv}f_{ji}^{S}-\frac{1}{2S_{i}-1}(ui|jv)$\\
			OV(1)-OV(1) & $\Delta A_{au,bv}=\frac{2}{2S_{i}-1}\delta_{uv}f_{ab}^{S}-\frac{1}{2S_{i}-1}(au|vb)$\\
			CV(1)-CO(1) & $\Delta A_{ai,vj}=(\sqrt{\frac{2S_{i}+1}{2S_{i}}}-1)(\delta_{ij}f_{av}^{\beta}-(av|ji))$\\
			CV(1)-OV(1) & $\Delta A_{ai,bv}=(\sqrt{\frac{2S_{i}+1}{2S_{i}}}-1)(-\delta_{ab}f_{vi}^{\alpha}-(ab|vi))$\\
			CO(1)-OV(1) & $\Delta A_{ui,bv}=\frac{1}{2S_{i}-1}((ui|vb)-(ub|vi))$\\
			OO(1)-OO(1) & $\Delta A_{tu,vw}=0$\\
			CV(1)-OO(1) & $\Delta A_{ai,vw}=-(\sqrt{\frac{2S_{i}+1}{2S_{i}-1}}-1)(av|wi)+\frac{1}{S_{i}}\sqrt{\frac{2S_{i}+1}{2S_{i}-1}}\delta_{vw}f_{ia}^{s}$\\
			CO(1)-OO(1) & $\Delta A_{ui,vw}=(\sqrt{\frac{2S_{i}}{2S_{i}-1}}-1)(-\delta_{uv}f_{iw}^{\alpha}-(uv|wi))+\frac{1}{\sqrt{2S_{i}(2S_{i}-1)}}\delta_{vw}f_{iu}^{\beta}$\\
			OV(1)-OO(1) & $\Delta A_{au,vw}=(\sqrt{\frac{2S_{i}}{2S_{i}-1}}-1)(\delta_{wu}f_{av}^{\beta}-(av|wu))-\frac{1}{\sqrt{2S_{i}(2S_{i}-1)}}\delta_{vw}f_{au}^{\alpha}$\\
			\hline
			\hline
		\end{tabular}
	\end{table*}

\subsection{Comparison with related works}\label{sec:comparison}
In this section, we compare different approaches for addressing spin contamination 
in SF-CIS and SF-TDA. A summary of the major formal differences among SA-SF-DFT, Q-SF-TDA, MRSF-TDA, and XSF-TDA is provided in Table~\ref{tab:comparison}.
Below we discuss the key distinctions among these methods are discussed in details.

\begin{table*}[htbp]
    \centering
    \caption{\raggedright Formal comparison between SA-SF-DFT\cite{zhang2015spin}, Q-SF-TDA\cite{chibueze2025spin}, MRSF-TDA\cite{lee2018eliminating}, and XSF-TDA.
    }
    \begin{tabular}{ccccc}
    \hline
    \hline
         method & reference & spin-adapted configuration & Hermitian & XC functional \\
         \hline
         SA-SF-DFT\cite{zhang2015spin} & any $S_i$ & all & no & collinear \\
         Q-SF-TDA\cite{chibueze2025spin} & any $S_i$ & OO(1) & yes & noncollinear \\
         MRSF-TDA\cite{lee2018eliminating} & $S_i=1$ & OO(1), CO(1), OV(1) & yes & collinear \\
         XSF-TDA & any $S_i$ & all & yes & noncollinear \\
         \hline
         \hline
    \end{tabular}
    \label{tab:comparison}
\end{table*}

SC-SF-CIS\cite{sears2003spin} include more determinants (for achieving spin completeness) than those necessary for spin-adaptation of SF-CIS. 
For instance, for the triplet model shown in Fig. \ref{fig:confs},
SC-SF-CIS includes all the 6 determinants for the CV type into the configuration space, which results in 2 spin-adapted configurations with spin zero. In comparison, TCIS or XSF-TDA employs a minimal set of spin-adapted configurations required to spin-adapt SF-CIS or SF-TDA.
To spin-adapt the CV($\alpha\beta$) type of excitation, only the following spin-adapted configuration is included
\begin{eqnarray}
&&[T_{ai}^\dagger(1)\times |S_i\rangle\rangle]^{S_i-1}_{S_i-1} \nonumber\\
&=& \sqrt{\frac{2S_{i}-1}{2S_{i}+1}}\left(|\Psi_{i}^{\bar a}\rangle
 -\frac{1}{2S_{i}}\sum_{t}(
      |\Psi_{it}^{a\bar t}\rangle -
      |\Psi_{\bar it}^{\bar a\bar t}\rangle)\right. \nonumber\\
&& \left.-\frac{1}{2S_{i}(2S_{i}-1)}\sum_{tu}|\Psi_{\bar iut}^{a\bar u\bar t}\rangle\right).
\end{eqnarray}
Additionally, an important difference between SC-SF-CIS and XSF-TDA is that the latter incorporates dynamical correlation, which is absent in the former.

SA-SF-DFT by Zhang and Herbert\cite{zhang2015spin} employs TEOM of the first kind\cite{li2010Spin} to derive its equations and uses an empirical matrix element in the form of collinear SF-TD-DFT to translate results
from SA-SF-CIS to SA-SF-DFT. Therefore, the dimension of SA-SF-CIS (or SA-SF-DFT)
is exactly the same as TCIS (or XSF-TDA). 
However, a notable difference between TEOM and TCIS in the spin-flip–down case ($S_f=S_i-1$)
is that TEOM yields a non-Hermitian generalized eigenvalue equation due to a nonvanishing coupling in the CV-OO block\cite{zhang2015spin}
(as $\langle 0|q_I q_J^\dagger \ne  \delta_{IJ}\langle 0|$ in this case), a point that has also been recognized in a recent work by Chibueze and Visscher\cite{chibueze2025spin}.
Therefore, even SA-SF-CIS itself is actually not a CI-type method. It is not equivalent to neither TCIS introduced in this work nor the SC-SF-CIS approach, a point that has also been realized in Ref. \cite{wang2020analytic}.
%However, XSF-TDA adopts the TCIS framework to derive the working equation following our earlier work for spin-conserving excitations\cite{li2011spin2}. 

Q-SF-TDA\cite{chibueze2025spin} recently introduced by Chibueze and Visscher\cite{chibueze2025spin} can be viewed as an approximation of XSF-TDA, which completely
neglects the spin-adaptation correction term in Eq. \eqref{eq:XSFTDA}. As a result, only the OO(1) type of excitations is spin-adapted in Q-SF-TDA, and other type of excitations are not corrected. This explains the very similar performance between Q-SF-TDA and USF-TDA observed in Ref. \cite{chibueze2025spin}.

MRSF-TDA by Lee et al.\cite{lee2018eliminating,lee2019efficient} follows a different strategy, utilizing a mixed-reference reduced density matrix that includes the $M_S=\pm1$ components of the
triplet ground state. Several key differences distinguish MRSF-TDA from XSF-TDA:
\begin{enumerate}
    \item MRSF-TDA is limited to triplet reference, while XSF-TDA is applicable to any high-spin reference.

    \item MRSF-TDA correctly describes the OO(1), CO(1), and
    OV(1) types of excitations in the triplet case, but
    does not correctly describe the spin symmetry of the CV($\alpha\beta$) type of excitations\cite{lee2018eliminating}. Because it only considers
    the additional contribution from the last determinant for the CV($\alpha\beta$) type in Fig. \ref{fig:confs}.
    
    \item We can recast the MRSF-TDA equation, see Eq. (S8.10) in Supporting Information of Ref. \cite{lee2018eliminating}, in the form of SF-TDA plus a correction term, which can be written according to the ordering of basis in Eq. \eqref{eq:OpBasis} (i.e., CV, CO, OV, and OO) as
\begin{eqnarray}
\left[\begin{array}{cccc}
-\mathbf{C}_{33} & -\mathbf{C}_{31} & -\mathbf{C}_{32} & \mathbf{0} \\ 
-\mathbf{C}_{13} & -\mathbf{C}_{11} & -\mathbf{C}_{12} & \mathbf{0} \\ 
-\mathbf{C}_{23} & -\mathbf{C}_{21} & -\mathbf{C}_{22} & \mathbf{0} \\ 
\mathbf{0} & \mathbf{0} & \mathbf{0} & \mathbf{0} \\ 
\end{array}\right],\label{eq:MRSFTDA}
\end{eqnarray}
with $(\mathbf{C}_{11})_{ui,wj} = 
c_\rX\langle\Psi^{M_S=+1}_{\bar{u}i}|\hat{H}|
\Psi^{M_S=-1}_{w\bar{j}}\rangle$ being
the introduced coupling between $\Psi^{M_S=+1}_{\bar{u}i}$
and $\Psi^{M_S=-1}_{w\bar{j}}$
originating from the spin-flip transitions from the
$M_S=+1$ and $M_S=-1$ components of the mixed reference
state\cite{lee2018eliminating}, respectively.
Then, it becomes evident that MRSF-TDA lacks corrections for couplings between OO-type excitations and other types. These can be essential for certain cases, see Sec. \ref{sec:INVEST}.

    \item Another practical difference is that to the best of our knowledge, the current MRSF-TDA
implementation in the GAMESS program\cite{GAMESS}
only supports collinear kernel,
whereas we use noncollinear kernel with ALDA0\cite{li2012theoretical}
in XSF-TDA, implemented using the PySCF package\cite{2020Recent,2018PySCF} in this work.
\end{enumerate}
Interestingly, we observe that for the triplet case, 
although derived from a different approach\cite{lee2018eliminating}, the final MRSF-TDA corrections for the CO(1) and OV(1) types of excitations are very similar to our correction terms. Specifically, consider the triplet case with two open-shell orbitals $u$ and $v$ shown in Fig. \ref{fig:confs}, then an example for the MRSF-TDA correction for the CO(1)-CO(1) block reads
\begin{eqnarray}
-(\mathbf{C}_{11})_{ui,uj} &=&
- c_\rX\langle\Psi^{M_S=+1}_{\bar{u}i}|\hat{H}|
\Psi^{M_S=-1}_{u\bar{j}}\rangle \nonumber\\
&=&
-c_\rX\langle v\bar{j}\|\bar{v}i\rangle
=
c_\rX(vi|jv),\label{eq:cXMRSFTDA}
\end{eqnarray}
while in our case, following the second line of Table \ref{tab:mats_dA}, we have 
\begin{align}
g_\rX \Delta A_{ui,uj} &=
g_\rX[2f_{ji}^S - (ui|ju)] \nonumber\\
&= g_\rX [\sum_{t=u,v} (jt|ti)-(ui|ju)]
= g_\rX (vi|jv),
\end{align}
which is similar to Eq. \eqref{eq:cXMRSFTDA} except for the prefactor.
It should be noted that the parameter $c_\rX$ used in MRSF-TDA \eqref{eq:cXMRSFTDA} and $g_\rX$ defined in Eq. \eqref{eq:rX} in general take different values. For the BHHLYP functional\cite{1988Density,1988Development,Becke1993Density}, which is one of the most frequently used XC functionals in MRSF-TDA, $c_\rX=0.5$ and $g_\rX=0.65$ if $g_{\rLDA}$ is set to 0.3 in the present work.

\section{Results}\label{sec:results}
\subsection{Benchmark for spin-flip transitions}
In this section, we examine how the excited states dominated by the OO-type of excitations are affected by the correction terms introduced in XSF-TDA. 
We first calculated the quartet excitation energies for 19 open-shell nonlinear molecules with doublet ground states
selected from the QUEST database\cite{Loos2025The} 
by using spin-flip-down TDA starting from the lowest quartet
excited state (denoted by D$\leftarrow$Q)
with the aug-cc-pVTZ basis\cite{1989Gaussian}. 
For comparison, we also performed calculations
using spin-flip-up TDA starting from the 
doublet ground state (denoted by D$\rightarrow$Q),
which is free of spin contamination with a ROKS reference\cite{li2010Spin}.
Linear diatomic molecules in the QUEST database are excluded, as either the ground doublet state or the lowest quartet state is spatially degenerate such that the ROKS calculations cannot be applied without breaking the spatial symmetry\cite{seth2005calculation}. 
Theoretical best estimates (TBEs) of vertical transition energies provided by the QUEST database were used as reference values. We used several XC functionals, including SVWN5\cite{1980Accurate}, BLYP\cite{1988Density,1988Development}, B3LYP\cite{Becke1993Density,1994Ab}, BHHLYP\cite{1988Density,1988Development,Becke1993Density}, and pure Hartree-Fock, which are characterized by increasing proportions of HF exchange.

\begin{figure}
    \centering
    \includegraphics[width=0.7\textwidth]{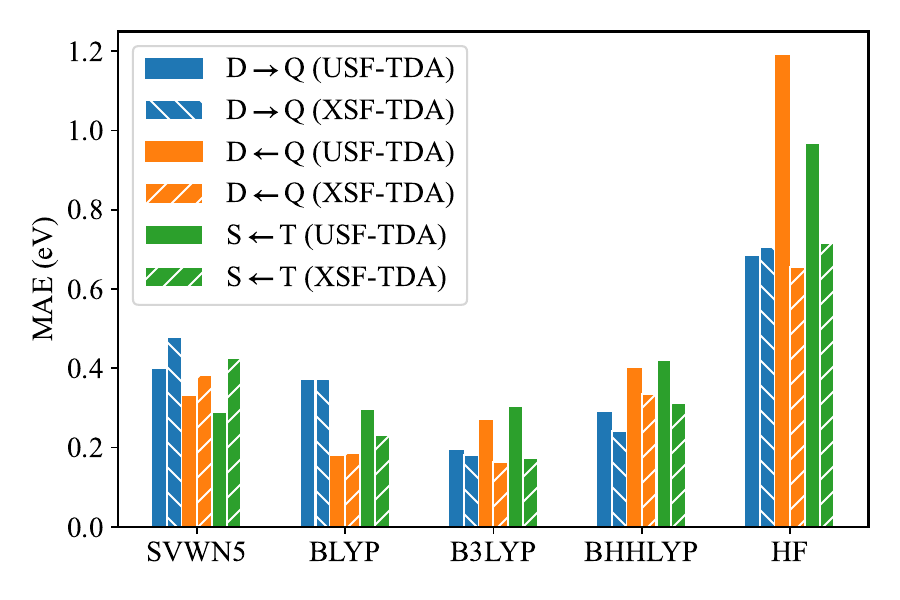}
    \caption{\raggedright Mean absolute errors (MAEs) in eV of vertical excitation energies computed with spin-flip TDA methods. D$\to$Q: spin-flip-up SF-TDA with UKS/ROKS reference; D$\leftarrow$Q and S$\leftarrow$T: spin-flip-down SF-TDA with UKS/ROKS reference. In spin-flip-down XSF-TDA method, Eq. \eqref{eq:XSFTDA} is used, while in spin-flip-up XSF-TDA, the analog of Eq. \eqref{eq:SFTDA} is used, as there is no spin contamination in such case.}
    \label{fig:qd-st}
\end{figure}

The detailed results obtained using USF-TDA and XSF-TDA methods are summarized in Tables S1 and S2 of the Supporting Information, and the mean absolute errors (MAEs) are displayed in Fig. \ref{fig:qd-st}. For USF-TDA, we also compute the change of the
expectation value of the square of the total spin angular momentum $\hat{S}^2$
following Ref. \cite{li2011spin}, as a diagnostic for spin contaminations.
The ideal value for spin-flip-down excitation is $\Delta\langle \hat{S}^2 \rangle = 
\langle \hat{S}^2 \rangle_{\mathrm{ex}} - \langle \hat{S}^2 \rangle_{\mathrm{gs}}=
(S_i-1)S_i - S_i(S_i+1) = -2S_i$. As shown in Table S2, the values of $\Delta\langle \hat{S}^2 \rangle$ obtained in USF-TDA only slightly deviate from the ideal value -3 for a quartet reference. Therefore, for these states, USF-TDA can 
deliver reasonably accurate results. As shown in Fig. \ref{fig:qd-st}, XSF-TDA with the same XC functional provides similar performances compared with USF-TDA for these excited states dominated by the OO type of excitations, indicating that introduced spin-adapted correction term does not deteriorate the performance, and hence the double counting of correlation is largely mitigated. To better illustrate this,
we performed calculations using XSF-TDA/SVWN5 with other $g_{\rLDA}$ values.
The MAE obtained without spin-adapted correction ($g_{\rLDA}=0.0$) is 0.36 eV,
while that obtained without scaling down the spin-adapted correction ($g_{\rLDA}=1.0$) is 0.44 eV. This indicates that the correction term ($g_{\rLDA}=1.0$) introduces double counting that worsens the description of
those excited states without spin contamination. Using a smaller value
$g_{\rLDA}=0.3$ gives a MAE of 0.38 eV, which is a reasonably good 
compromise.

% \begin{figure}
%     \centering
%     \includegraphics[width=1.0\linewidth]{figures/mae_qd.pdf}
%     \caption{Mean absolute errors (MAEs) in eV of vertical excitation energies computed in XSF-TDA with quartet ROKS reference. $g_{\rLDA}$ is the parameters in Eq.\ref{eq:rX}}
%     \label{fig:glda}
% \end{figure}

% \begin{table}[!ht]
%     \centering
%     \caption{Mean absolute errors (MAEs) in eV of vertical excitation energies computed with quartet ROKS reference. $g_{\rLDA}$ is the parameters in Eq.\ref{eq:rX}}
%     \begin{tabular}{ccccc}
%     \hline
%     \hline
%         ~ & SVWN5 & BLYP & B3LYP & BHHLYP \\ 
%         \hline
%         $g_{\rLDA}$=0.0 & 0.36  & 0.17  & 0.21  & 0.34  \\
%         $g_{\rLDA}$=0.3 & 0.38  & 0.19  & 0.16  & 0.34  \\ 
%         $g_{\rLDA}$=1.0 & 0.44  & 0.24  & 0.25  & 0.33  \\
%         \hline
%         \hline
%     \end{tabular}
%     \label{table:glda}
% \end{table}

Apart from calculations with a quartet reference, we also computed the excitation energies from the lowest triplet to the singlet ground state (denoted by S$\leftarrow$T) for 89 molecules from the QUEST database using USF-TDA and XSF-TDA
with a triplet reference. The detailed results are provided in Table S3. We find that the performances of XSF-TDA for systems with different initial spins are largely consistent,
with the difference in MAE being within 0.05 eV.
For SVWN5, the MAE of XSF-TDA is slightly larger than that of USF-TDA by 0.14 eV. However, the MAEs of XSF-TDA 
for BLYP, B3LYP, BHHLYP, and HF are smaller than those of USF-TDA.

\subsection{Be and Mg}\label{sec:BeAndMg}
We next examine the performance of XSF-TDA for excited states involving other types of excitations.
We calculated the excitation energies of the $^3P_z$ and $^1P_{x,y,z}$  states for \ce{Be} and \ce{Mg} with
the $^1S$ ground state, using the USF-TDA and XSF-TDA methods starting from the $^3P_z$ reference state.
The involved excitations for $^3P_z$, $^1P_z$, $^1P_{x,y}$ are of O$_1$O$_2$(1), O$_1$O$_1$(1), and OV(1) types.
All calculations were performed under the $D_{2h}$ point group symmetry using the 6-31G\cite{1980Self} and aug-cc-pVTZ basis sets. The computed results are summarized in Table \ref{tab:be-mg}.

As indicated by the $\Delta \langle \hat{S}^2\rangle$ values, the $^3P_z$ and $^1P_z$ states are almost free of
spin contamination in USF-TDA, for which XSF-TDA give very similar results. In constrast, the $^1P_{x,y}$ states of Be and Mg are severely spin contaminated in USF-TDA. This contamination leads to a substantial reduction in the excitation energy of these states relative to the $^1P_z$ state, while they are expected to be triply degenerate. 
For example, the splittings between the $^1P_{x,y}$ and $^1P_z$ states computed by USF-TDA with BHHLYP and the aug-cc-pVTZ basis set are 1.2 eV and 0.7 eV for Be and Mg, respectively, while XSF-TDA produces a markedly smaller splittings of only 0.10 eV and 0.25 eV. In general, XSF-TDA consistently produces much smaller energy differences between the $^1P_z$ and $^1P_{x,y}$ states of Be and Mg
than USF-TDA  with all the tested XC functionals and basis sets. Across all error metrics, including mean error (ME), MAE, standard deviation (SE), and maximum error (MAX), XSF-TDA demonstrates superior performance over USF-TDA. 

\begin{table}[!ht]
\setlength{\tabcolsep}{0.1pt}
    \centering
    \caption{\raggedright Excited states for \ce{Be} and \ce{Mg} obtained using USF-TDA and XSF-TDA with different basis sets. The data in parentheses indicate the values of $\Delta\langle \hat{S}^{2}\rangle$ in USF-TDA calculation. The first excited state $^3 P_z$ is taken as the reference. ME: mean error, MAE: mean absolute error, SE: standard error, MAX: maximum error.}\label{tab:be-mg}
    \footnotesize
    \begin{tabular}{ccccccccccccccc}
    \hline
    \hline
        ~ & ~ & ~ & ~ & ~ & ~ & USF-TDA & ~ & ~ & ~ & ~ & ~ & XSF-TDA & ~ & ~    \\ \cmidrule(r){5-9} \cmidrule(r){11-15}
        Basis & Atom & State & Expt.\cite{2013NIST} & SVWN5 & BLYP & B3LYP & BHHLYP & HF & ~ & SVWN5 & BLYP & B3LYP & BHHLYP & HF  \\ \hline
        6-31G & \ce{Be} & $^3P_z$ & 2.73  & 2.20  & 2.17  & 2.33  & 2.60  & 2.11  & ~ & 2.20  & 2.17  & 2.35  & 2.59  & 2.13  \\ 
        ~ & ~ & ~ & ~ & (-2.00) & (-2.00) & (-2.00) & (-2.00) & (-2.00) & ~ & ~ & ~ & ~ & ~ &   \\ 
        ~ & ~ & $^1P_{x,y}$ & 5.28  & 3.88  & 4.01  & 3.94  & 3.79  & 4.09  & ~ & 4.46  & 4.61  & 4.81  & 5.10  & 5.97    \\ 
        ~ & ~ & ~ & ~ & (-1.00) & (-1.00) & (-1.00) & (-1.00) & (-1.00) & ~ & ~ & ~ & ~ & ~ &    \\ 
        ~ & ~ & $^1P_z$ & 5.28  & 4.52  & 4.82  & 4.91  & 5.06  & 6.04  & ~ & 4.50  & 4.79  & 4.87  & 5.03  & 5.98    \\ 
        ~ & ~ & ~ & ~ & (-1.99) & (-1.98) & (-1.99) & (-1.99) & (-1.99) & ~ & ~ & ~ & ~ & ~ & ~   \\ 
        ~ & \ce{Mg} & $^3P_z$ & 2.71  & 2.66  & 2.86  & 2.86  & 2.94  & 2.13  & ~ & 2.68  & 2.87  & 2.87  & 2.96  & 2.17    \\ 
        ~ & ~ & ~ & ~ & (-2.00) & (-2.00) & (-2.00) & (-2.00) & (-1.99) & ~ & ~ & ~ & ~ & ~ &    \\ 
        ~ & ~ & $^1P_{x,y}$ & 4.35  & 3.63  & 3.88  & 3.79  & 3.70  & 3.46  & ~ & 4.03  & 4.29  & 4.38  & 4.57  & 4.62    \\ 
        ~ & ~ & ~ & ~ & (-1.00) & (-1.00) & (-1.00) & (-1.00) & (-1.00) & ~ & ~ & ~ & ~ & ~ &    \\ 
        ~ & ~ & $^1P_z$ & 4.35  & 3.96  & 4.23  & 4.28  & 4.43  & 4.72  & ~ & 3.94  & 4.18  & 4.22  & 4.35  & 4.54    \\ 
        ~ & ~ & ~ & ~ & (-1.98) & (-1.95) & (-1.96) & (-1.97) & (-1.93) & ~ & ~ & ~ & ~ & ~ & ~   \\ 
        % ~ & ~ & ~ & ~ & ~ & ~ & ~ & ~ & ~ & ~ & ~ & ~ & ~ & ~    \\ 
        aug-cc-pVTZ & \ce{Be} & $^3P_z$ & 2.73  & 2.15  & 2.25  & 2.45  & 2.58  & 2.06  & ~ & 2.16  & 2.17  & 2.35  & 2.58  & 2.11   \\ 
        ~ & ~ & ~ & ~ & (-1.99) & (-2.00) & (-2.00) & (-2.00) & (-1.99) & ~ & ~ & ~ & ~ & ~    \\ 
        ~ & ~ & $^1P_{x,y}$ & 5.28  & 3.64  & 3.63  & 3.63  & 3.58  & 3.82  & ~ & 4.10  & 4.12  & 4.35  & 4.63  & 5.21     \\ 
        ~ & ~ & ~ & ~ & (-1.00) & (-1.00) & (-1.00) & (-1.00) & (-1.00) & ~ & ~ & ~ & ~ & ~   \\ 
        ~ & ~ & $^1P_z$ & 5.28  & 4.17  & 4.37  & 4.55  & 4.78  & 5.61  & ~ & 4.11  & 4.24  & 4.39  & 4.53  & 5.20   \\ 
        ~ & ~ & ~ & ~ & (-1.93) & (-1.93) & (-1.92) & (-1.93) & (-1.83) & ~ & ~ & ~ & ~ & ~     \\ 
        ~ & \ce{Mg} & $^3P_z$ & 2.71  & 2.67  & 2.91  & 2.87  & 2.92  & 2.06  & ~ & 2.69  & 2.92  & 2.89  & 2.96  & 2.14     \\ 
        ~ & ~ & ~ & ~ & (-2.00) & (-2.00) & (-2.00) & (-2.00) & (-1.99) & ~ & ~ & ~ & ~ & ~     \\ 
        ~ & ~ & $^1P_{x,y}$ & 4.35  & 3.59  & 3.82  & 3.73  & 3.64  & 3.30  & ~ & 3.93  & 4.15  & 4.23  & 4.38  & 4.19     \\ 
        ~ & ~ & ~ & ~ & (-1.00) & (-1.00) & (-1.00) & (-1.00) & (-1.00) & ~ & ~ & ~ & ~ & ~    \\ 
        ~ & ~ & $^1P_z$ & 4.35  & 3.87  & 4.12  & 4.19  & 4.34  & 4.50  & ~ & 3.83  & 4.01  & 4.05  & 4.14  & 4.12     \\ 
        ~ & ~ & ~ & ~ & (-1.95) & (-1.89) & (-1.91) & (-1.91) & (-1.74) & ~ & ~ & ~ & ~ & ~     \\ 
        ~ & ME & ~ & ~ & -0.71  & -0.53  & -0.49  & -0.42  & -0.46  & ~ & -0.56  & -0.41  & -0.30  & -0.13  & -0.09     \\ 
        ~ & MAE & ~ & ~ & 0.71  & 0.59  & 0.54  & 0.51  & 0.73  & ~ & 0.56  & 0.47  & 0.36  & 0.26  & 0.39     \\ 
        ~ & SE & ~ & ~ & 0.47  & 0.52  & 0.53  & 0.60  & 0.67  & ~ & 0.36  & 0.41  & 0.34  & 0.31  & 0.45     \\ 
        ~ & MAX & ~ & ~ & -1.64  & -1.65  & -1.65  & -1.70  & -1.46  & ~ & 1.18  & 1.16  & 0.93  & 0.75  & 0.70     \\ \hline \hline
    \end{tabular}
\end{table}

\subsection{Potential energy curves for \ce{HF}}
\begin{figure}[htbp]
\centering
\begin{tabular}{cc}
   \includegraphics[width=0.49\textwidth]{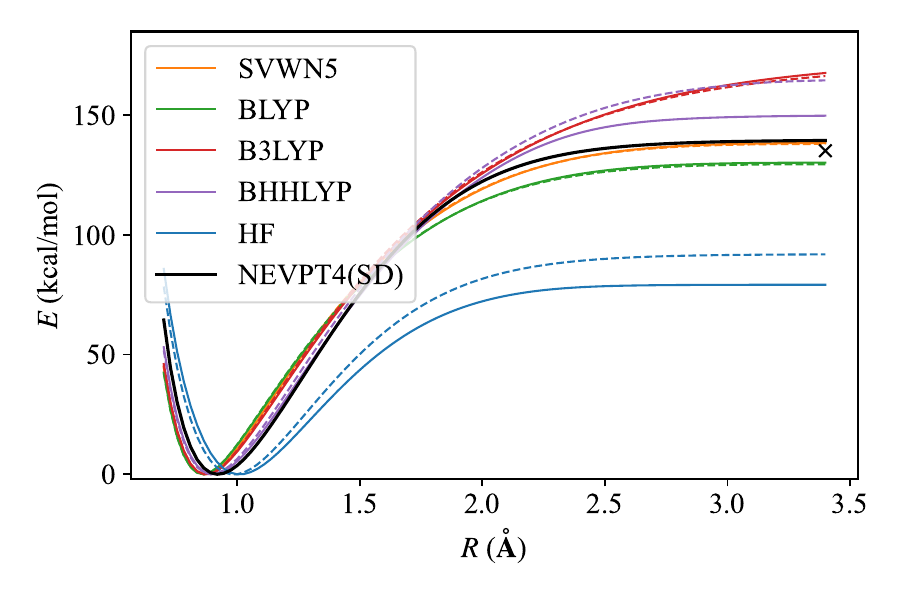}  & 
   \includegraphics[width=0.49\textwidth]{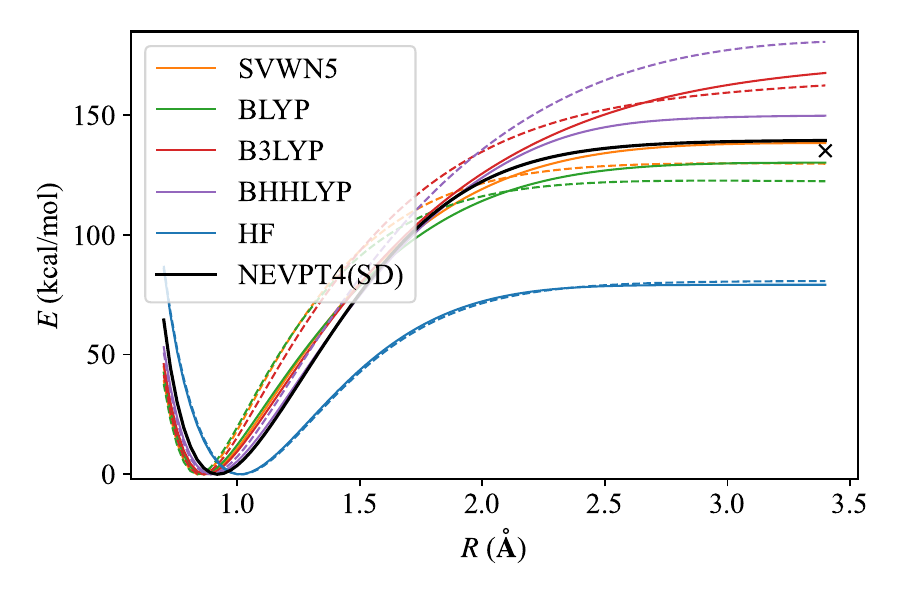} \\
   (a) USF-TDA vs. XSF-TDA & (b) USF-TDA vs. MRSF-TDA \\
   \includegraphics[width=0.49\textwidth]{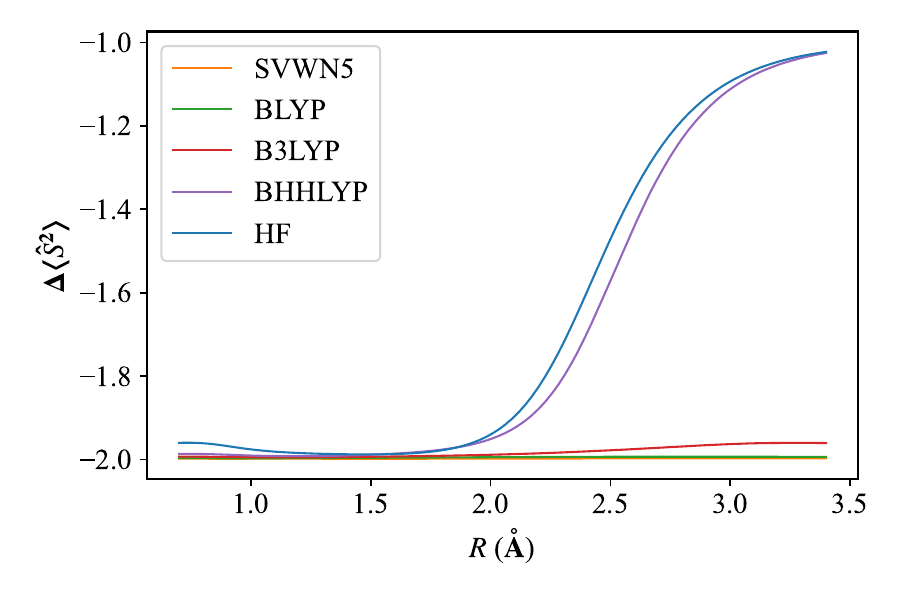} & 
   \includegraphics[width=0.49\textwidth]{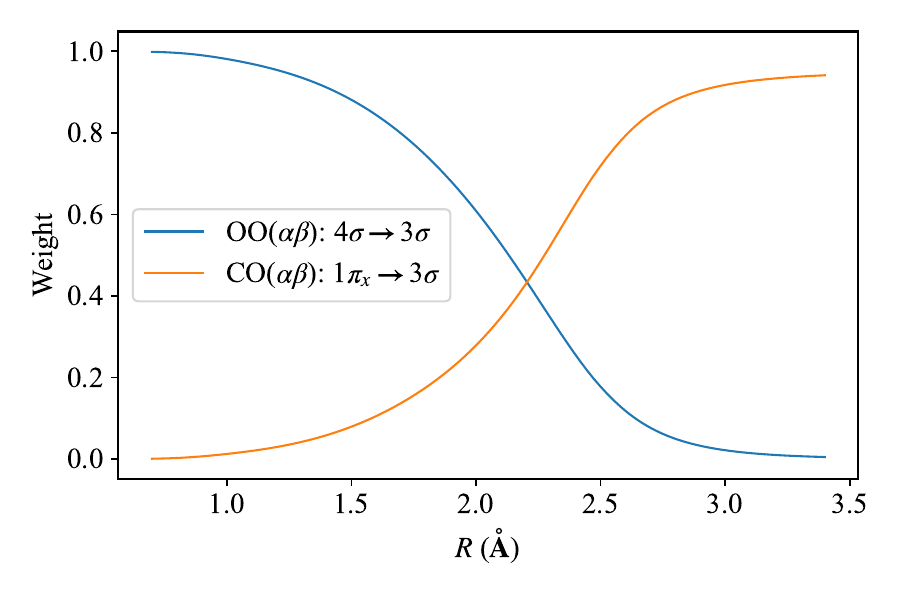} \\
   (c) $\Delta\langle\hat{S}^2\rangle$ in USF-TDA & (d) Weight obtained in USF-TDA/BHHLYP
\end{tabular}
    \caption{\raggedright 
    Results for the \ce{HF} molecule using SF-TDA methods with the aug-cc-pVTZ basis set. (a) PECs obtained with
    USF-TDA (solid) and XSF-TDA (dashed);
    (b) PECs obtained with USF-TDA (solid) and MRSF-TDA (dashed); 
    (c) $\Delta\langle \hat{S}^{2}\rangle$ obtained in USF-TDA;
    (d) Weights of the dominant two excitations obtained in USF-TDA/BHHLYP. The minimum of each curve in (a) and (b) has been shifted to zero for a better comparison. The cross marker in (a) and (b) represents
    the experimental bond dissociation energy 135.1 kcal/mol\cite{1970Bond}.}
    \label{fig:hf_pec}
\end{figure}

Apart from the excitation energies due to spin-flip transitions, we also applied XSF-TDA for computing
the potential energy curve (PEC) of \ce{HF},
which is a typical testbed for spin-flip methods\cite{sears2003spin,shao2003spin}.
We take the same reference state $^3\Sigma$
with the electronic configuration $1\sigma^{2}2\sigma^{2}1\pi^{4}3\sigma^{1}4\sigma^{1}$
as in previous works\cite{sears2003spin} to compute the PEC of the ground state using SF-TDA.
The PECs obtained using USF-TDA and XSF-TDA with the aug-cc-pVTZ basis set are shown in Fig. \ref{fig:hf_pec}(a).
%To obtain the correct reference state, we specify the electron occupation in specific irreducible representations in PySCF with $C_{2v}$ symmetry, such as A$_1$(4,2) B$_1$(1,1) B$_2$(1,1), which means that four alpha electrons and two beta electrons occupy the irreducible representation of A$_1$ and so on. 
For comparison, we also show the PECs obtained with MRSF-TDA\cite{lee2018eliminating} in Fig. \ref{fig:hf_pec}(b). 
To make a fair comparison, we make sure that the 
same reference state is obtained, by passing the converged
results from PySCF as initial guess to GAMESS
through MOKIT\cite{zou2022molecular} and using
the maximum overlap method (MOM)\cite{2008Self} 
in ROKS calculations. In Figs. \ref{fig:hf_pec}(a) and \ref{fig:hf_pec}(b), the PEC obtained from NEVPT4(SD) (fully internally contracted partial fourth-order $N$-electron valence perturbation theory)\cite{2025nevpt4} with a CAS(8e,5o) active space using the ORCA package \cite{neese2020orca,neese2022software} is shown as a reference.

As shown in Fig. \ref{fig:hf_pec}(a), the PECs obtained by XSF-TDA agree well with those obtained by USF-TDA for SVWN5,
BLYP and B3LYP, while the results differ for BHHLYP and
pure HF at large bond distances. To better understand such behaviors, we plot $\Delta\langle \hat{S}^{2}\rangle$ obtained in USF-TDA in Fig. \ref{fig:hf_pec}(c). It is clear that at large bond distance, the results obtained with BHHLYP and pure HF show a significant amount of spin contamination, which in turn leads to the growing discrepancies between the XSF-TDA and USF-TDA results
observed in Fig. \ref{fig:hf_pec}(a). The increase of $\Delta\langle \hat{S}^{2}\rangle$ can be explained by the increase of CO($\alpha\beta$) type of excitations as demonstrated in Fig. \ref{fig:hf_pec}(d). At shorter bond lengths, the ground state is primarily reached by OO($\alpha\beta$)-type excitations. As the bond length increases, the dominant configuration gradually changes to a CO($\alpha\beta$)-type excitation.

In comparison, we find that the MRSF-TDA result with pure HF is very close to the USF-TDA result, see Fig. \ref{fig:hf_pec}(b), likely due to the missing of the coupling between the OO and other types of excitations in the correction \eqref{eq:MRSFTDA}.
The major difference between the PECs of MRSF-TDA 
and those of USF-TDA with other XC functionals can
be attributed to the use of collinear functionals in the former case. In particular, for SVWN5 and BLYP, the correction term in MRSF-TDA reduce to zero, causing the results to coincide with those of collinear SF-TDA,
while the USF-TDA results are obtained with the noncollinear kernel \eqref{eq:SFkernel}.

\subsection{Potential energy curves for \ce{F2}}
\begin{figure}[htbp]
\centering
\begin{tabular}{c}
   \includegraphics[width=0.48\textwidth]{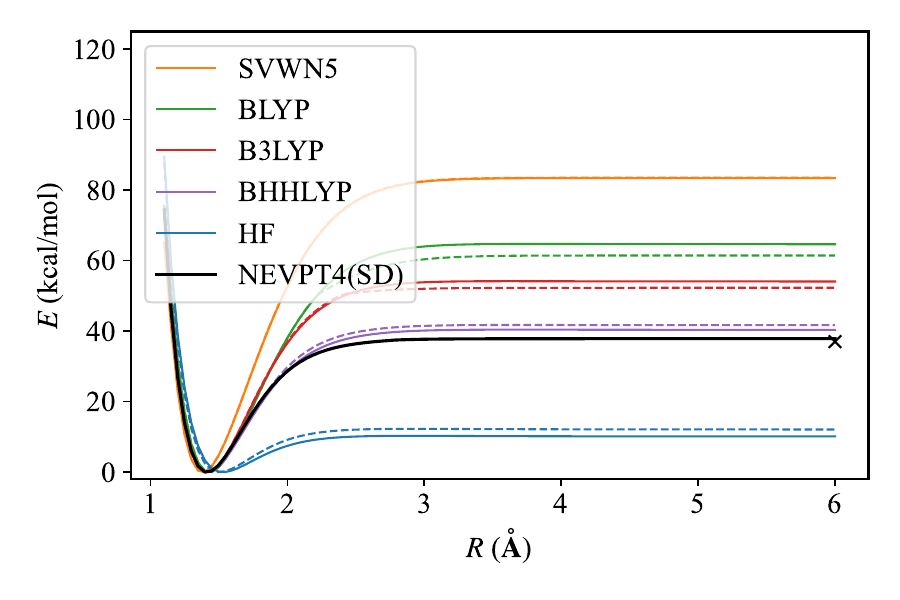}  
   \\
   (a) USF-TDA vs. XSF-TDA \\
   \includegraphics[width=0.48\textwidth]{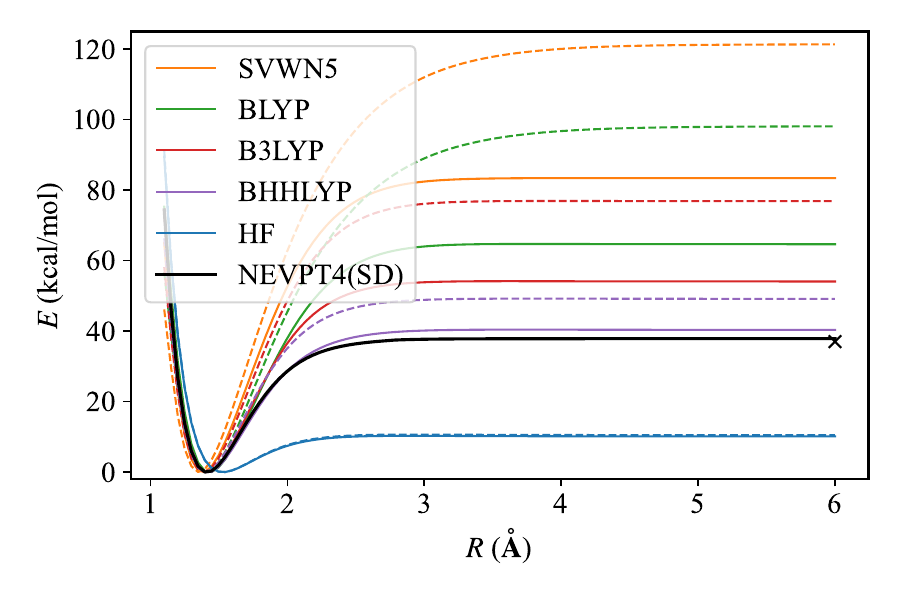} \\
   (b) USF-TDA vs. MRSF-TDA \\
   \includegraphics[width=0.48\textwidth]{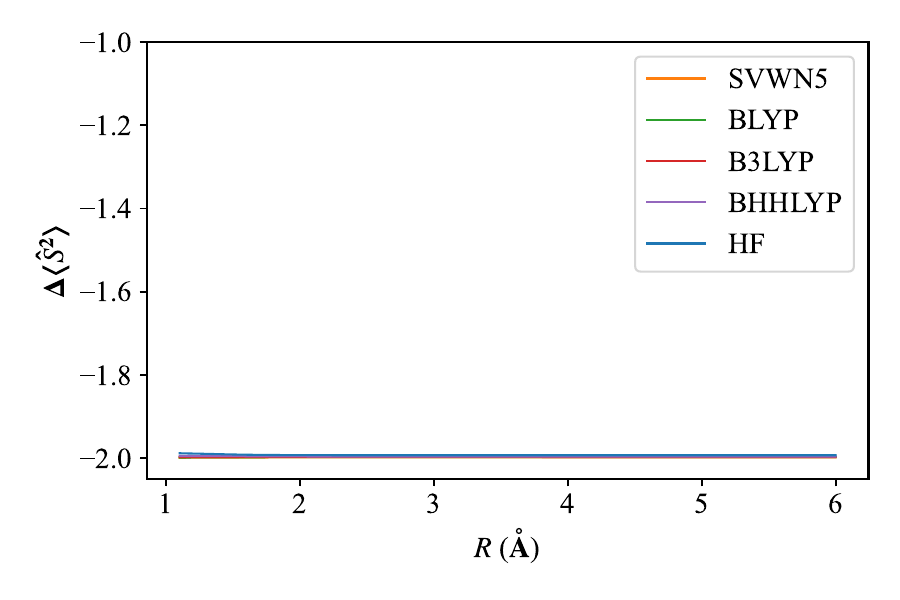}\\
   (c) $\Delta \langle \hat{S}^2\rangle$ in USF-TDA
\end{tabular}
    \caption{\raggedright 
    Results for the \ce{F2} molecule using SF-TDA methods
    with the aug-cc-pVTZ basis set. (a) PECs
    obtained with USF-TDA (solid) and XSF-TDA (dashed);
    (b) PECs obtained with USF-TDA (solid) and MRSF-TDA (dashed); (c) $\Delta \langle \hat{S}^2\rangle$ in USF-TDA.
    The minimum of each curve in (a) and (b) has been shifted to zero for a better comparison. The cross marker in (a) and (b) represents
    the experimental bond dissociation energy 37.0 kcal/mol\cite{2024Bond}.}
    \label{fig:f2_pec}
\end{figure}

The \ce{F2} molecule is a challenging system for electronic structure methods due to the coexistence of strong dynamical and nondynamical correlation effects. We investigated
its PEC using SF-TDA methods with the aug-cc-pVTZ basis set.
from the high-spin $^3\Sigma^{+}_{u}$ reference state
with the electronic configuration $1\sigma_{g}^{2}1\sigma_{u}^{2}2\sigma_{g}^{2}2\sigma_{u}^{2}3\sigma^{1}_{g}1\pi^{4}_{u}1\pi^{4}_{g}3\sigma_{u}^{1}$.
The resulting PECs obtained from USF-TDA, XSF-TDA, and MRSF-TDA are shown in Figs. \ref{fig:f2_pec}(a) and \ref{fig:f2_pec}(b), along with the NEVPT4(SD) results using a CAS(14e,8o) active space. For this system, we find that the PECs
obtained by USF-TDA and XSF-TDA are qualitatively similar for all the considered XC functionals. This is explained by the negligible amount of spin contamination in the USF-TDA results shown in Fig. \ref{fig:f2_pec}(c). For pure HF, the MRSF-TDA
result using a ROKS reference almost coincides with the USF-TDA result, which is also observed in the calculations of the \ce{HF} molecule.
However, the PECs of MRSF-TDA with other XC functionals tend to be higher than the corresponding USF-TDA results at larger bond distance, which is likely attributed to the use of collinear functionals. 

\subsection{Inverted singlet-triplet gap systems}\label{sec:INVEST}
\begin{figure*}
    \centering
    \begin{tabular}{c}
    \includegraphics[width=1.0\textwidth]{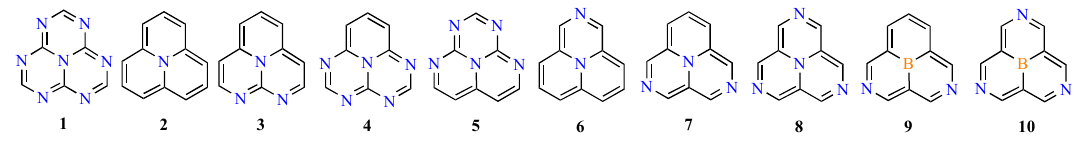} \\
    (a) Schematic representation of the benchmark systems\cite{2023heptazine} \\
    \includegraphics[width=1.0\textwidth]{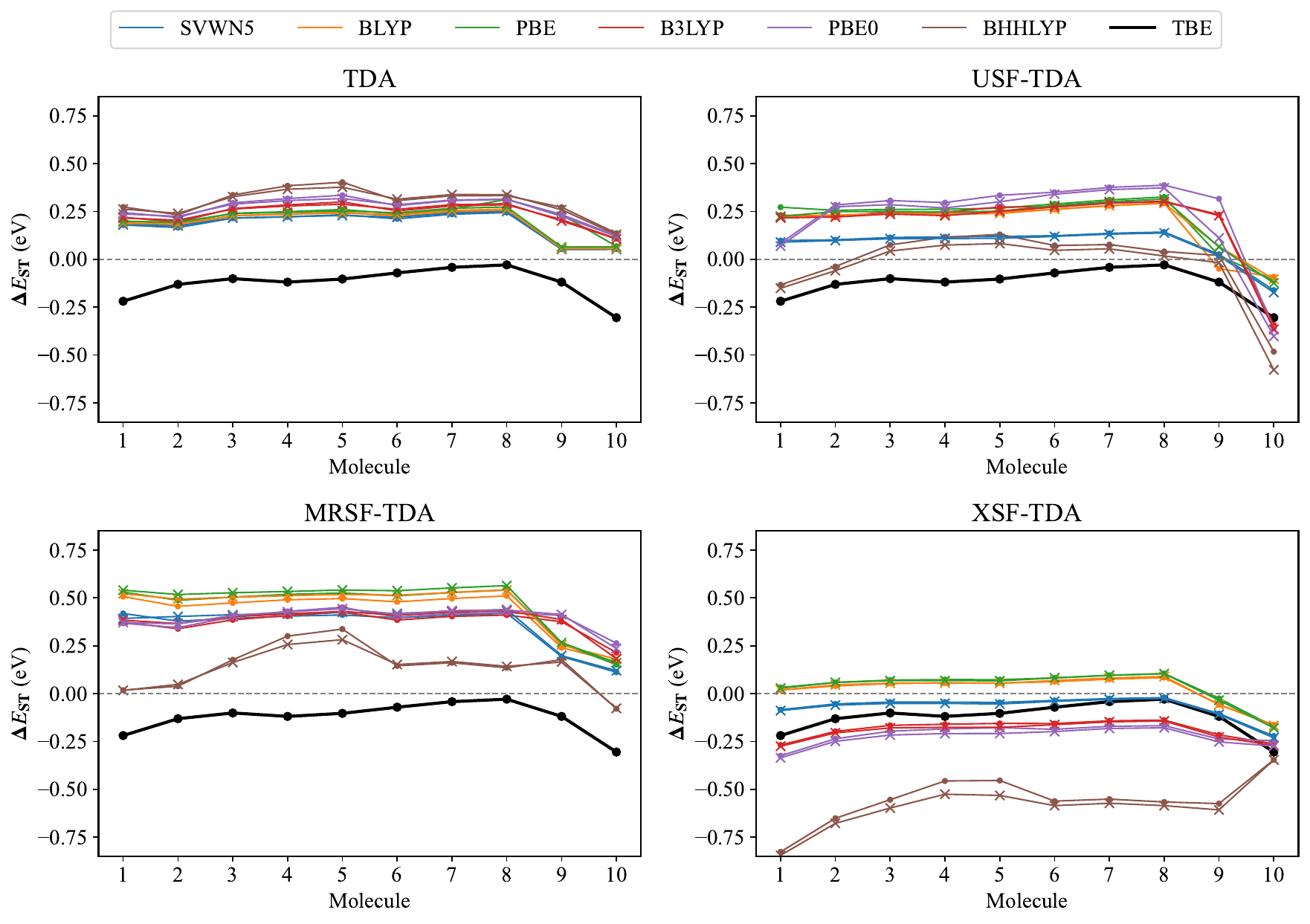} \\
    (b) Singlet-triplet gaps $\Delta E_{\mathrm{ST}}$ 
    \end{tabular}
    \caption{\raggedright 
    Results for inverted single-triplet gap systems.
    (a) Schematic representation of heptazine (no. 1) and cyclazine (no. 2) and their derivatives considered in Ref. \cite{2023heptazine} as benchmark systems.
    (b) Singlet-triplet gaps $\Delta E_{\mathrm{ST}}$ (in eV) computed using
    TDA and SF-TDA methods with the 6-31G* (cross) and aug-cc-pVTZ (dot) basis sets. The theoretical best estimates (TBEs) are taken from
    Ref. \cite{2023heptazine}.
    }
    \label{fig:INVEST}
\end{figure*}

The inverted singlet-triplet gap (abbreviated as INVEST) systems, in which the Hund's rule is violated for the lowest singlet (S$_1$) and triplet (T$_1$)
excited states, have recently gained a lot of  interests\cite{2019Singlet,jpclett.9b02333,Ricci2021Singlet,2021Organic,0Large,jpca.1c10492,0Delayed,D2TC02508F,D2CP02364D,2023Connections,2023The,2023Symmetry,fchem.2023.1239604,drwal2023role,lashkaripour2025addressing}
due to their potential in optelectronic applications such as 
organic light-emitting diodes (OLED). Unfortunately, standard
spin-conserving TD-DFT cannot correctly describe the correct
ordering of S$_1$ and T$_1$\cite{jpclett.9b02333}, except for some doubly hybrid functionals\cite{jpca.1c10492}. Here, we apply spin-conserving and spin-flip TDA methods to the benchmark set by Loos et al.\cite{2023heptazine},
see Fig. \ref{fig:INVEST}(a), which contains 10 molecules with
negative $\Delta E_{\mathrm{ST}}=E(\mathrm{S}_1)-E(\mathrm{T}_1)<0$.
The molecular coordinates were optimized at the level of CCSD(T)/cc-pVTZ (within the frozen-core approximation), and TBEs for $\Delta E_{\mathrm{ST}}$ were obtained from CC3/aug-cc-pVTZ + [CCSDT/6-31+G(d) -  CC3/6-31+G(d)] for S$_1$ and CC3/aug-cc-pVDZ + [CCSD/aug-cc-pVTZ - CCSD/aug-cc-pVDZ] for T$_1$, respectively. For spin-conserving TDA, $\Delta E_{\mathrm{ST}}$ is obtained from
two separate TDA calculations for S$_1$ and T$_1$, respectively, starting from the closed-shell ground state S$_0$,
while for spin-flip-down TDA, $\Delta E_{\mathrm{ST}}$ is taken as the
excitation energy of the second singlet excited state $\omega_1$
with respect to the triplet reference T$_1$. [Note that the excitation energy of the first singlet excited state $\omega_0$ is the energy difference between $S_0$ and $T_1$.] Fig. \ref{fig:INVEST}(b) displays the obtained $\Delta E_{\mathrm{ST}}$ by spin-conserving TDA, USF-TDA, MRSF-TDA,
XSF-TDA with the 6-31G* and aug-cc-pVTZ basis sets using
the following XC functionals: SVWN5\cite{1980Accurate}, BLYP\cite{1988Development,1988Density}, PBE\cite{1998Generalized}, B3LYP\cite{Becke1993Density,1994Ab}, PBE0\cite{1999Assessment,1999Toward}, and BHHLYP\cite{1988Density,1988Development,Becke1993Density}.
For both MRSF-TDA and XSF-TDA, the ROKS reference is used and the stability of the solution has been checked. For the molecules 9 and 10, some solutions obtained
with pure density functionals were found unstable, and in such case we used the occupation obtained with in BHHLYP calculations to obtain a stable solution.
Detailed results are summarized in Table S4-S7. 

As shown in Fig. \ref{fig:INVEST}(b), the obtained
$\Delta E_{\mathrm{ST}}$ in both spin-conserving and spin-flip-down TDA
show a weak dependence on the basis set.
Conventional spin-conserving TDA gives all positive $\Delta E_{\mathrm{ST}}$,
agree with previous studies\cite{jpclett.9b02333,jpca.1c10492}.
USF-TDA with BHHLYP can give negative $\Delta E_{\mathrm{ST}}$ for a few systems, while other XC functionals can only give negative values for the last two systems. Notably, while the excited state S$_1$ is dominated by the OO($\alpha\beta$) type of excitations, we find that results of USF-TDA with
hybrid functionals display substantial amount of spin contamination (see Tables S4 and S6). For MRSF-TDA, most of $\Delta E_{\mathrm{ST}}$ are also positive. Special double tuning of the parameters in XC functionals
is required to achieve good performance, as demonstrated in a recent study\cite{lashkaripour2025addressing}. [NB: the energy of T$_1$ in Ref. \cite{lashkaripour2025addressing} is taken as the energy obtained in a second MRSF-TDA calculation for the lowest triplet. In Table S8, we made a comparison of results 
obtained by these two approaches.] 
Quite remarkably, XSF-TDA with B3LYP and PBE0 show very good performances,
with MAEs being 0.07 eV and 0.10 eV, respectively, for results using the aug-cc-pVTZ basis set (see Table S7). We attribute the better performance of XSF-TDA over other spin-flip TDA methods to the presence of the coupling between the OO(1) and other type of excitations in the spin-adapted corrections, see the last three lines in Table \ref{tab:mats_dA}. In contrast, such couplings are missing in MRSF-TDA, see Eq. \eqref{eq:MRSFTDA}. 
In particular, the coupling between the OO(1) and CV(1) type of excitations
can increase the percentage of CV(1) type of excitations in the S$_1$ state,
which is responsible for the dynamical spin polarization\cite{kollmar1978violation,drwal2023role,lashkaripour2025addressing}
that is essential for obtaining negative $\Delta E_{\mathrm{ST}}$.
Besides, the too negative $\Delta E_{\mathrm{ST}}$ obtained from XSF-TDA/BHHLYP suggest that it is worthwhile to systematically explore the choice of $g_{\rX}$ 
in future work.

% \begin{table}[!ht]
%     \centering
%     \caption{Singlet-triplet gaps $\Delta E_{\mathrm{ST}}$ (in eV) computed using
%     MRSF-TDA and XSF-TDA methods with the 6-31G* in BHHLYP functional. The XSF-TDA0 method represents the $g_{\rX}$ parameter in Eq.\ref{eq:rX} using 0.5 and does not consider the contribution of the OO($\alpha\beta$) terms, and with the collinear XC kernel.  ME: mean error, MAE: mean absolute error.}
%     \begin{tabular}{ccccc}
%     \hline
%     \hline
%         ~ & TBE & MRSF-TDA & XSF-TDA0 & XSF-TDA \\ 
%         \hline
%         1 & -0.22  & 0.02  & 0.05  & -0.84  \\ 
%         2 & -0.13  & 0.05  & 0.07  & -0.68  \\ 
%         3 & -0.10  & 0.16  & 0.18  & -0.60  \\ 
%         4 & -0.12  & 0.26  & 0.27  & -0.53  \\ 
%         5 & -0.10  & 0.28  & 0.30  & -0.53  \\ 
%         6 & -0.07  & 0.15  & 0.17  & -0.59  \\ 
%         7 & -0.04  & 0.17  & 0.19  & -0.57  \\ 
%         8 & -0.03  & 0.14  & 0.17  & -0.59  \\ 
%         9 & -0.12  & 0.17  & 0.18  & -0.61  \\ 
%         10 & -0.30  & -0.08  & -0.04  & -0.35  \\ 
%         ~ & ME & 0.26  & 0.28  & -0.46  \\ 
%         ~ & MAE & 0.26  & 0.28  & 0.46  \\ 
%         \hline
%         \hline
%     \end{tabular}
%     \label{table:mrsf}
% \end{table}

\section{Conclusion}
In this work, we introduced a simple and accurate method for resolving the spin contamination problem in SF-TDA,
extending on our previous work for spin-conserving excitations\cite{li2010Spin,li2011spin,li2011spin2}. Benchmark calculations demonstrate that the proposed XSF-TDA offers comparable accuracy as USF-TDA for excited states without significant amount of spin contamination, but dramatically improves upon USF-TDA for heavily spin contaminated excited states.
We have also made both formal and numerical between XSF-TDA and MRSF-TDA\cite{lee2018eliminating}, demonstrating the superiority of XSF-TDA in dealing with complex electronic structure problems such as the inverted singlet-triplet gap systems\cite{jpca.1c10492}.

This work opens up many promising directions to be explored in future: (1) While XSF-TDA with the simple parameterization in Eq. \eqref{eq:rX} adopted in this work has been demonstrated
to be superior to USF-TDA, there is still room for improving its performance by systematic optimization of $g_\rX$ against a broader benchmark.
(2) The combination of multicollinear functionals\cite{li2023noncollinear,zhang2025spin,wang2025zero} with XSF-TDA is straightforward and further comparison with
ALDA0\cite{li2012theoretical} will be made.
(3) Implementing spin-orbit couplings among different spin states\cite{wang2005simplified,li2013combining,chibueze2024restricted} and nonadiabatic couplings between excited states\cite{li2014first,li2014first2,zhang2014analytic} 
will enable the study of nonadiabatic excited-state dynamics
for open-shell systems. Progress along these lines will be reported in due time.

\section*{Acknowledgment}
This work is dedicated to Prof. Wenjian Liu on the occasion of his 60th birthday.
The authors acknowledge Zikuan Wang, Jingxiang Zou, Bohan Zhang for critically reading the manuscript. This work was supported by the Quantum Science and Technology-National Science and Technology Major Project(2023ZD0300200) and the Fundamental Research Funds for the Central Universities.

%\section{References}

\bibliographystyle{tfo}
\bibliography{main}

\begin{thebibliography}{80}
\providecommand{\url}[1]{\texttt{#1}}
\providecommand{\urlprefix}{URL }

\bibitem{shao2003spin}
Y. Shao, M. Head-Gordon and A.I. Krylov,  J. Chem. Phys.  \textbf{118} (11),
  4807--4818 (2003).

\bibitem{casanova2020spin}
D. Casanova and A.I. Krylov,  Phys. Chem. Chem. Phys.  \textbf{22} (8),
  4326--4342 (2020).

\bibitem{kitzmann2022spin}
W.R. Kitzmann, J. Moll and K. Heinze,  Photochem. Photobiol. Sci.  \textbf{21}
  (7), 1309--1331 (2022).

\bibitem{kitzmann2023charge}
W.R. Kitzmann and K. Heinze,  Angew. Chem. Int. Ed  \textbf{62} (15),
  e202213207 (2023).

\bibitem{zhang2015spin}
X. Zhang and J.M. Herbert,  J. Chem. Phys.  \textbf{143} (23), 234107 (2015).

\bibitem{2019Analytic}
J. Mato and M.S. Gordon,  J. Phys. Chem. A  \textbf{123} (6), 1260--1272
  (2019).

\bibitem{Keipert2014Dynamics}
Y. Harabuchi, K. Keipert, F. Zahariev, T. Taketsugu and M.S. Gordon,  J. Phys.
  Chem. A  \textbf{118} (51), 11987--11998 (2014).

\bibitem{herbert2022spin}
J.M. Herbert and A. Mandal, in \emph{Time-dependent density functional theory}
  (Jenny Stanford Publishing, Singapore, 2022), pp. 361--404.

\bibitem{wang2004time}
F. Wang and T. Ziegler,  J. Chem. Phys.  \textbf{121} (24), 12191--12196
  (2004).

\bibitem{wang2005performance}
F. Wang and T. Ziegler,  J. Chem. Phys.  \textbf{122} (7), 074109 (2005).

\bibitem{wang2003comparison}
F. Wang and W. Liu,  J. Chin. Chem. Soc.  \textbf{50} (3B), 597--606 (2003).

\bibitem{gao2004time}
J. Gao, W. Liu, B. Song and C. Liu,  J. Chem. Phys.  \textbf{121} (14),
  6658--6666 (2004).

\bibitem{gao2005time}
J. Gao, W. Zou, W. Liu, Y. Xiao, D. Peng, B. Song and C. Liu,  J. Chem. Phys.
  \textbf{123} (5), 054102 (2005).

\bibitem{rinkevicius2010spin}
Z. Rinkevicius, O. Vahtras and H. {\AA}gren,  J. Chem. Phys.  \textbf{133}
  (11), 114104 (2010).

\bibitem{bernard2012general}
Y.A. Bernard, Y. Shao and A.I. Krylov,  J. Chem. Phys.  \textbf{136} (20),
  204103 (2012).

\bibitem{li2012theoretical}
Z. Li and W. Liu,  J. Chem. Phys.  \textbf{136} (2), 024107 (2012).

\bibitem{li2016critical}
Z. Li and W. Liu,  J. Chem. Theor. Comput.  \textbf{12} (6), 2517--2527 (2016).

\bibitem{li2023noncollinear}
H. Li, Z. Pu, Q. Sun, Y.Q. Gao and Y. Xiao,  J. Chem. Theor. Comput.
  \textbf{19} (8), 2270--2281 (2023).

\bibitem{zhang2025spin}
X. Zhang and Y. Xiao,  arXiv preprint arXiv:2505.17766   (2025).

\bibitem{wang2025zero}
T. Wang, H. Li, Y.Q. Gao and Y. Xiao,  J. Chem. Theor. Comput.  \textbf{21}
  (14), 6905--6921 (2025).

\bibitem{casida2005propagator}
M.E. Casida,  J. Chem. Phys.  \textbf{122} (5), 054111 (2005).

\bibitem{li2010Spin}
Z. Li and W. Liu,  J. Chem. Phys.  \textbf{133} (6), 157 (2010).

\bibitem{sears2003spin}
J.S. Sears, C.D. Sherrill and A.I. Krylov,  J. Chem. Phys.  \textbf{118} (20),
  9084--9094 (2003).

\bibitem{vahtras2007general}
O. Vahtras and Z. Rinkevicius,  J. Chem. Phys.  \textbf{126} (11), 114101
  (2007).

\bibitem{rowe1975tensor}
D. Rowe and C. Ngo-Trong,  Rev. Mod. Phys.  \textbf{47} (2), 471 (1975).

\bibitem{li2011spin}
Z. Li, W. Liu, Y. Zhang and B. Suo,  J. Chem. Phys.  \textbf{134} (13), 134101
  (2011).

\bibitem{li2011spin2}
Z. Li and W. Liu,  J. Chem. Phys  \textbf{135} (19), 194106 (2011).

\bibitem{li2016criticalDD}
Z. Li and W. Liu,  J. Chem. Theor. Comput.  \textbf{12} (1), 238--260 (2016).

\bibitem{chibueze2025spin}
C.S. Chibueze and L. Visscher,  J. Chem. Phys.  \textbf{163} (9), 094111
  (2025).

\bibitem{lee2018eliminating}
S. Lee, M. Filatov, S. Lee and C.H. Choi,  J. Chem. Phys.  \textbf{149} (10),
  104101 (2018).

\bibitem{lee2019efficient}
S. Lee, E.E. Kim, H. Nakata, S. Lee and C.H. Choi,  J. Chem. Phys.
  \textbf{150} (18), 184111 (2019).

\bibitem{1981Second}
P. J{\o}rgensen, \emph{Second quantization-based methods in quantum chemistry}
   (Elsevier, Amsterdam, Netherlands, 2012).

\bibitem{1979Spin}
R. Pauncz, \emph{Spin Eigenfunctions}   (Springer US, New York, 1979).

\bibitem{wang2020analytic}
Z. Wang, Z. Li, Y. Zhang and W. Liu,  J. Chem. Phys.  \textbf{153} (16) (2020).

\bibitem{GAMESS}
G.M.J. Barca, C. Bertoni, L. Carrington, D. Datta, N. De~Silva, J.E. Deustua,
  D.G. Fedorov, J.R. Gour, A.O. Gunina, E. Guidez, T. Harville, S. Irle, J.
  Ivanic, K. Kowalski, S.S. Leang, H. Li, W. Li, J.J. Lutz, I. Magoulas, J.
  Mato, V. Mironov, H. Nakata, B.Q. Pham, P. Piecuch, D. Poole, S.R. Pruitt,
  A.P. Rendell, L.B. Roskop, K. Ruedenberg, T. Sattasathuchana, M.W. Schmidt,
  J. Shen, L. Slipchenko, M. Sosonkina, V. Sundriyal, A. Tiwari, J.L.
  Galvez~Vallejo, B. Westheimer, M. Wloch, P. Xu, F. Zahariev and M.S. Gordon,
  J. Chem. Phys.  \textbf{152} (15), 154102 (2020).

\bibitem{2020Recent}
Q. Sun, X. Zhang, S. Banerjee, P. Bao, M. Barbry, N.S. Blunt, N.A. Bogdanov,
  G.H. Booth, J. Chen, Z.H. Cui {\em{et~al.}},  J. Chem. phys.  \textbf{153}
  (2), 024109 (2020).

\bibitem{2018PySCF}
Q. Sun, T.C. Berkelbach, N. Blunt, G.H. Booth, S. Guo, Z. Li, J. Liu, J.D.
  Mcclain, E. Sayfutyarova and S. Sharma,  Wires. Comput. Mol. Sci  \textbf{8}
  (1), e1340 (2018).

\bibitem{1988Density}
A.D.P. Becke,  Phys. Rev. A  \textbf{38} (6), 3098--3100 (1988).

\bibitem{1988Development}
C. Lee, W. Yang and R.G. Parr,  Phys. Rev. B  \textbf{37} (2), 785--789 (1988).

\bibitem{Becke1993Density}
A.D. Becke,  J. Chem. Phys.  \textbf{98} (7), 5648--5652 (1993).

\bibitem{Loos2025The}
P.F. Loos, M. Boggio-Pasqua, A. Blondel, F. Lipparini and D. Jacquemin,  J.
  Chem. Theor. Comput.  \textbf{21} (16), 8010--8033 (2025).

\bibitem{1989Gaussian}
T.H. Dunning,  J. Chem. Phys.  \textbf{90}, 1007--1023 (1989).

\bibitem{seth2005calculation}
M. Seth and T. Ziegler,  J. Chem. Phys.  \textbf{123} (14), 144105 (2005).

\bibitem{1980Accurate}
S.H. Vosko, L. Wilk and M. Nusair,  Can. J. Phys.  \textbf{58} (8), 1200--1211
  (1980).

\bibitem{1994Ab}
P.J. Stephens, F.J. Devlin, C.F. Chabalowski and M.J. Frisch,  J. Phys. Chem.
  \textbf{98} (1-3), 247--257 (1994).

\bibitem{1980Self}
R. Krishnan, J.S. Binkley, R. Seeger and J.A. Pople,  J. Chem. Phys.
  \textbf{72} (1), 650--654 (1980).

\bibitem{2013NIST}
A. Kramida, Y. Ralchenko, J. Reader and N.A. Team, NIST Atomic Spectra Database
  (version 5.1) [online]  2013.

\bibitem{zou2022molecular}
J. Zou, Molecular Orbital Kit (MOKIT), \url{https://gitlab.com/jxzou/mokit}
  2025, (accessed Jun 25, 2025).

\bibitem{2008Self}
A.T.B. Gilbert, N.A. Besley and P.M.W. Gill,  J. Phys. Chem. A  \textbf{112}
  (50), 13164--71 (2008).

\bibitem{2025nevpt4}
E.M. Kempfer, K. Sivalingam and F. Neese,  J. Chem. Theor. Comput.  \textbf{21}
  (8), 3953--3967 (2025).

\bibitem{neese2020orca}
F. Neese, F. Wennmohs, U. Becker and C. Riplinger,  J. Chem. Phys.
  \textbf{152} (22), 224108 (2020).

\bibitem{neese2022software}
F. Neese,  WIRES Comput. Molec. Sci.  \textbf{12} (5), e1606 (2022).

\bibitem{1970Bond}
B. De~Darwent,  NSRDS-NBS 31   (1970).

\bibitem{2024Bond}
P. Wang, S. Gong and Y. Mo,  J. Phys. Chem. Lett.  \textbf{15} (51),
  12594--12600 (2024).

\bibitem{2019Singlet}
J. Ehrmaier, E.J. Rabe, S.R. Pristash, K.L. Corp and W. Domcke,  J. Phys. Chem.
  A  \textbf{123} (38), 8099--8108 (2019).

\bibitem{jpclett.9b02333}
P. de~Silva,  J. Phys. Chem. Lett.  \textbf{10} (18), 5674--5679 (2019).

\bibitem{Ricci2021Singlet}
G. Ricci, E. San-Fabi{\'a}n, Y. Olivier and J.C. Sancho-Garc{\'\i}a,  Chem.
  Phys. Chem.  \textbf{22} (6), 553--560 (2021).

\bibitem{2021Organic}
R. Pollice, P. Friederich, C. Lavigne, G.D.P. Gomes and A. Aspuru-Guzik,
  Matter  \textbf{4} (5), 1654--1682 (2021).

\bibitem{0Large}
F. Dinkelbach, M. Bracker, M. Kleinschmidt and C.M. Marian,  J. Phys. Chem. A
  \textbf{125} (46), 10044--10051 (2021).

\bibitem{jpca.1c10492}
S. Ghosh and K. Bhattacharyya,  J. Phys. Chem. A  \textbf{126} (8), 1378--1385
  (2022).

\bibitem{0Delayed}
N. Aizawa, Y.J. Pu, Y. Harabuchi, A. Nihonyanagi, R. Ibuka, H. Inuzuka, B.
  Dhara, Y. Koyama, K.i. Nakayama, S. Maeda {\em{et~al.}},  Nature
  \textbf{609} (7927), 502--506 (2022).

\bibitem{D2TC02508F}
G. Ricci, J.C. Sancho-García and Y. Olivier,  J. Mater. Chem. C  \textbf{10},
  12680--12698 (2022).

\bibitem{D2CP02364D}
L. Tučková, M. Straka, R.R. Valiev and D. Sundholm,  Phys. Chem. Chem. Phys.
  \textbf{24}, 18713--18721 (2022).

\bibitem{2023Connections}
E. Monino and P.F. Loos,  J. Chem. Phys.  \textbf{159} (3), 034105 (2023).

\bibitem{2023The}
D. Drwal, M. Matousek, P. Golub, A. Tucholska, M. Hapka, J. Brabec, L. Veis and
  K. Pernal,  J. Chem. Theor. Comput.  \textbf{19} (21), 7606--7616 (2023).

\bibitem{2023Symmetry}
J.T. Blaskovits, M.H. Garner and C. Corminboeuf,  Angew. Chem. Int. Ed
  \textbf{62}, e202218156 (2023).

\bibitem{fchem.2023.1239604}
A. Dreuw and M. Hoffmann,  Front. Chem.  \textbf{11}, 1239604 (2023).

\bibitem{drwal2023role}
D. Drwal, M. Matousek, P. Golub, A. Tucholska, M. Hapka, J. Brabec, L. Veis and
  K. Pernal,  J. Chem. Theory Comput.  \textbf{19} (21), 7606--7616 (2023).

\bibitem{lashkaripour2025addressing}
A. Lashkaripour, W. Park, M. Mazaherifar and C.H. Choi,  J. Chem. Theory
  Comput.  \textbf{21} (11), 5661--5668 (2025).

\bibitem{2023heptazine}
P.F. Loos, F. Lipparini and D. Jacquemin,  J. Phys. Chem. Lett.  \textbf{14}
  (49), 11069--11075 (2023).

\bibitem{1998Generalized}
J.P. Perdew, K. Burke and M. Ernzerhof,  Phys. Rev. Lett.  \textbf{77} (18),
  3865--3868 (1998).

\bibitem{1999Assessment}
M. Ernzerhof and G.E. Scuseria,  J. Chem. Phys.  \textbf{110} (11), 5029--5036
  (1999).

\bibitem{1999Toward}
C. Adamo and V. Barone,  J. Chem. Phys.  \textbf{110} (13), 6158--6170 (1999).

\bibitem{kollmar1978violation}
H. Kollmar and V. Staemmler,  Theor. Chim. Acta  \textbf{48} (3), 223--239
  (1978).

\bibitem{wang2005simplified}
F. Wang and T. Ziegler,  J. Chem. Phys.  \textbf{123} (15), 154102 (2005).

\bibitem{li2013combining}
Z. Li, B. Suo, Y. Zhang, Y. Xiao and W. Liu,  Mol. Phys.  \textbf{111} (24),
  3741--3755 (2013).

\bibitem{chibueze2024restricted}
C.S. Chibueze and L. Visscher,  J. Chem. Phys.  \textbf{161} (9), 094112
  (2024).

\bibitem{li2014first}
Z. Li and W. Liu,  J. Chem. Phys.  \textbf{141} (1), 014110 (2014).

\bibitem{li2014first2}
Z. Li, B. Suo and W. Liu,  J. Chem. Phys.  \textbf{141} (24), 244105 (2014).

\bibitem{zhang2014analytic}
X. Zhang and J.M. Herbert,  J. Chem. Phys.  \textbf{141} (6), 064104 (2014).

\end{thebibliography}

\end{document}